\documentclass[useAMS,usenatbib]{mn2e}
\usepackage{epsfig}
\usepackage{textcomp}
\newcommand{\ltsima} {$\; \buildrel < \over \sim \;$}
\newcommand{\gtsima} {$\; \buildrel > \over \sim \;$}
\newcommand{\lta} {\lower.5ex\hbox{\ltsima}}
\newcommand{\gta} {\lower.5ex\hbox{\gtsima}}

\title[Synchrotron emission \& Outflow dynamics ]{Synchrotron emission in GRBs observed by Fermi: Its limitations and the role of the photosphere.}
\author[S. Iyyani et al.]{S. Iyyani$^{1,2,3,4}$\thanks{email: shabuiyyani@particle.kth.se},  
 F. Ryde$^{1,2}$,
J.M. Burgess$^{1,2}$,
A. Pe'er$^{5}$,  
D. B{\' e}gu{\'e}$^{1,2}$\\
$^{1}$Department of Physics, KTH Royal Institute of Technology, AlbaNova, SE-106 91 Stockholm, Sweden\\ 
$^{2}$The Oskar Klein Centre for Cosmoparticle Physics, AlbaNova, SE-106 91 Stockholm, Sweden\\ 
$^{3}$Department of Physics, Stockholm University, AlbaNova, SE-106 91 Stockholm, Sweden\\ 
$^{4}$Erasmus Mundus Joint Doctorate in Relativistic Astrophysics\\
$^{5}$Department of Physics, University Cork, 5 Brighton Villas, Western Road, Cork, Ireland\\
}

\begin{document}

\date{Accepted... Received...; in original form ...}

\pagerange{\pageref{firstpage}--\pageref{lastpage}} \pubyear{2015}

\maketitle

\label{firstpage}

\begin{abstract}

It has been suggested that the prompt emission in gamma-ray bursts consists of several components giving rise to the observed spectral shape. Here we examine a sample of {  the 8 brightest,  single} pulsed {\it Fermi} bursts  whose spectra are modelled by using synchrotron emission as one of the components. Five of these bursts require an additional photospheric component (blackbody). %{\it , while the other three are consistent with %a single synchrotron emission alone.}
%component.  
In particular, we investigate the inferred properties of the jet and the physical requirements set by the observed components {  for these five bursts, in the context of a baryonic dominated outflow, motivated by the strong  {photospheric} component.} 
We find similar jet properties for {  all five} bursts:  the bulk Lorentz factor  decreases monotonously over the pulses and lies between 1000 and 100. This evolution is robust  and can neither be explained by a varying radiative efficiency nor a varying magnetisation of the jet assuming the photosphere radius is above the coasting radius). Such a behaviour challenges several dissipation mechanisms, e.g., the internal shocks. %%%%%%%Furthermore, the flow nozzle, at which the flow starts to accelerate, varies between $10^6$ and $10^9$ cm. 
Furthermore, in all 8 cases the data clearly reject a fast-cooled synchrotron spectrum {  (in which a significant fraction of the emitting electrons have cooled to energies below the minimum injection energy), }
%%%%%%%%%%that the electrons should not have lost most of their energy during the emission episode. 
 % i.e., they are  not radiating efficiently (in order to maintain the observed electron distribution) 
{  inferring} a typical electron Lorentz factor of  $10^4 - 10^7$. Such values are much higher than what is typically expected in internal shocks.   Therefore, while the synchrotron scenario is not rejected by the data, the interpretation does present several limitations that need to be addressed. 
%These results require that only a small fraction of the electrons at the dissipation site are accelerated and that the emission is radiatively inefficient. Alternatively, continuous reheating of the electrons can be invoked instead, in order to maintain the observed electron distribution, relaxing the requirement on the $\gamma_{\rm min}$% that is  required to fit the data 
%as well as increasing the radiative efficiency.
Finally, we point out and discuss alternative interpretations.% of the data, such as subphotospheric dissipation models, {  Poynting flux dominated models}, and models invoking external shocks.}

\end{abstract}

\section{Introduction}  

%Increasing detections have been made of blackbody component in Gamma ray bursts (GRBs) spectra \citep{Guiriec2011,Axelsson2012}. Photospheric emission is naturally expected within a standard fireball scenario (e.g. \cite{Meszaros2006}) where the emission is a blackbody provided there is no sub photospheric dissipation.
%According to a standard fireball scenario  (e.g. \cite{Meszaros2006}),  
%the optically thick plasma formed at the base of the outflow expands, thereby converting its internal energy to the kinetic energy of the plasma till the saturation radius, $r_s$ is reached. At some point the optical depth falls to unity from where the photons escape the plasma and this surface is referred to as the photosphere. 
%High above the photosphere the kinetic energy of the outflow gets dissipated by some mechanisms like internal shocks resulting in high energy electrons which thereafter cool by radiating via non-thermal optically thin synchrotron emission process resulting in the non-thermal component of the spectrum. 

Since gamma-ray burst (GRB) spectra mostly have a non-thermal shape, an early suggestion for the emission mechanism was optically-thin synchrotron emission \citep{Katz1994,Tavani1996,Rees1994,Sari1998}.  The viability of this model has been mainly ascertained by studying the low-energy photon index, $\alpha$, of the Band function fits (e.g. \cite{Preece1998,Goldstein2013}). A large fraction of bursts have an $\alpha> -2/3$ which is incompatible with the simplest models of synchrotron emission. %The 'line of death' of synchrotron emission has been marked at $\alpha = -2/3$ (slow cooled synchrotron) (Goldstein et al. 2013). %However, recently Burgess et al. 2014 showed that a slow cooled synchrotron emission spectrum when modelled using the Band function gives an $\alpha = -0.8$ . This actually worsens the 'line of death' issue of synchrotron emission as now there is a larger sample of GRB spectra that are inconsistent with the synchrotron emission. 
In addition to this, \cite{Axelsson2015} and {  Yu et al.  (2015)} studied the width of the $\nu F_{\nu}$ spectrum %of the Band function fits for  $\sim 1900$ bursts. It was 
and found that a majority of  long GRBs are too narrow to be explained by synchrotron emission, even from the most narrow electron distributions. The bursts with the narrowest spectra are even consistent 
%Moreover, there have also been bursts detected by {\it Fermi} gamma ray space telescope which are consistent 
with a single Planck function through out the burst duration \citep{Ryde2004, Ghirlanda2013, Larsson2015}.

A possible explanation to these observations was given by the two-emission--zone model, which combines photospheric emission (quasi-Planck spectrum) with a non-thermal component  \citep{Meszaros&Rees2000,Meszaros2002}. The latter  component is expected to be emitted from dissipation events in the optically-thin region of the flow. 
 Indeed, fits with a blackbody in combination with a non-thermal component did perform well  in many bursts \citep{Ryde2005, Ryde&Pe'er2009, Guiriec2011, Axelsson2012, Iyyani2013}. %(Axelsson et al. 2012; Guiriec et al. 2011, Burgess et al. 2014a). 
%Photospheric emission models are either single emission zone model or two emission zone model. In a single emission zone model, the entire GRB spectra is interpreted to have been emitted from the photosphere. The deviation of the observed spectrum from a pure Planck function is explained by either subphtospheric dissipation \citep{Rees&Meszaros2005,Ryde2010}%(Rees \& Meszaros 2005;Ryde et al. 2010) 
%or geometrical broadening \citep{Pe'er2008, Lundman2013}. 
 In these fits the non-thermal emission was modelled either by a power-law or a Band function.
%Most of the GRB spectra are fit by the Band function alone 
However, these are empirical functions and do not incorporate the actual emission physics. Therefore, they  can, at best,  only model the shape of physical spectra over a limited energy range. %This issue ascertains the significance of fitting physical models to GRB spectra. 

{  Initial steps fitting physical models using proper spectral deconvolution were made by \cite{Liang&Jernigan1983} and \cite{Tavani1996}.}  Similarly, \cite{Burgess2014a} studied the GRB spectra of a sample of the {  8 brightest} single-pulsed bursts by fitting a blackbody + synchrotron emission model. {  In these fits, the synchrotron component was calculated assuming a prescribed electron energy distribution.} %(see further \S \ref{sec:alt}}).
A significant blackbody component was identified in 5 of these bursts.  For three of the eight bursts  synchrotron emission alone is consistent with the data. However, for all cases a fast-cooling synchrotron emission is much too broad and is strongly contradicted by the data. Such spectra are expected when the cooling is complete and the electrons radiate most of their energy. The synchrotron emission that is permitted by the data {  indicate that a  majority of electrons, at the minimum injection energy, have not had time to cool significantly. %($\delta \gamma_{  min}/ \gamma_{  min} \ll 1$). 
In other words,} such emission can be denoted as incompletely cooled synchrotron emission, since it  could be due to {  one of several reasons:}
{  first}, the electrons might not have had time to loose most of their energy (so called slow-cooling; \cite{Sari1998, Asano&Terasawa2009, Zhang&Yan2011}), {  second,} reheating of the electrons can  compensate for the cooling \citep{Kumar&McMahon2008, Beniamini&Piran2014},  third, the electrons might be in a moderately fast cooling regime as described in \citet{Uhm&Zhang2014}. In such a model while the electrons are in the fast cooling regime,  $t_{\rm cool} < \sim 
t_{\rm dyn}$ and therefore the emergent spectrum is intermediate between slow and fast cooling,  {  and  fourth, if the  emission region has a varying magnetic field, only a fraction of the electrons will be able to cool efficiently, leaving a predominantly uncooled electron distribution \citep{Pe'er&Zhang2006, Beniamini&Piran2014, Zhang2015}.} %In comparison to the model, blackbody + Band function, it was found that blackbody when was added in combination with the synchrotron function produced more defined pulses and spectral evolution with time, which were distinct from that observed for the non-thermal emission. It was argued that this  suggests that the two spectral components originate from two different emission zones.  

In the current paper, we use the identified photospheric component of the {five} bursts to determine the properties of the flow at the photosphere {  (assuming it to be baryonic-dominated)}, such as Lorentz factor, $\Gamma$, photospheric radius, $r_{\rm ph}$, nozzle radius, $r_0$ and saturation radius, $r_s$ (\S \ref{outflow_cal}). Assuming that these properties are the same at the optically-thin emission site we study the synchrotron component to constrain the magnetic field strength, $B$ and the electron Lorentz factor, $\gamma_{\rm el}$ at the dissipation site (\S \ref{sync}). In \S 4 we investigate how and if a varying radiative efficiency or magnetisation can influence the determined parameter evolutions. We discuss the limitations of the presented interpretation in \S 5 and finally conclude in \S 6.

% in the 8 bursts within the sample of \cite{Burgess2014a}, in particularly, 5 among them which have a significant detection of blackbody.% and constrain the physical properties of the emission site (photosphere and optically thin dissipation site). Thus,
 %Using the observed spectral properties of the blackbody component, we estimate the outflow dynamics of the jet such as Lorentz factor, $\Gamma$, photospheric radius, $r_{\rm ph}$, nozzle radius, $r_0$ and saturation radius, $r_s$, and use them along with the synchrotron component fit values to constrain the magnetic field strength, $B$ and the electron Lorentz factor, $\gamma_{\rm el}$ at the dissipation site .
 % In \cite{Iyyani2013}, it was shown that GRB110721A possess varying jet properties: decreasing Lorentz factor, gradually increasing photospheric radius and a dramatically evolving nozzle radius of the jet. Here, we assess the temporal outflow properties of the jet of each burst and find that the deductions made by \cite{Iyyani2013} are indeed a common behaviour over individual pulse GRBs. 
%By studying the Lorentz factor, $\Gamma$, nozzle radius of the jet, $r_0$, photospheric radius, $r_{ph}$ and the saturation radius, $r_s$, of the outflow evolution with time, and also their evolution with respect to the observed spectral properties such observed luminosity, peak of non-thermal emission, observed total flux,  we propose a physical picture of how the outflow evolves with time in the bursts.  

\subsection{Sample and spectral properties}
\label{Sample}
%{  Reminder to the spectral analysis results of the bursts that were reported in Burgess et al. 2013.  }
%Here we would briefly describe the techniques that were followed for the analysis, \cite{Burgess2013a}. 
The eight bursts in the sample were selected by requiring that peak flux should be greater than 5 photons $\rm s^{-1} \, cm^{-2}$ in the energy range 10 keV to 40 MeV and that the light curves of the bursts should be single-peaked, in order to avoid overlap with different emission episodes. 
%However, we cannot be sure if there is any weaker emission episodes underlying the observed main peak of the burst. 
The bursts were binned following the Bayesian-block method \citep{Scargle2013}, which ensures that the binning is mainly determined by significant changes in the count rate. 
The following bursts were found to have a significant and strong blackbody component: GRB081224A \citep{Wilson-Hodge2008}, GRB090719A \citep{VanderHorst2009}, GRB100707A \citep{Wilson-Hodge2010}, GRB110920A \citep{McGlynn2012,Iyyani2015,Shenoy2013}, weak but statistically significant: 
GRB110721A \citep{Tierney2011},
while GRB081110A, GRB090809A and GRB110407A were found to be consistent with a synchrotron component alone.    

%Spectral analysis of these {\it Fermi} detected bursts, were found to be considerably consistent with slow cooling synchrotron emission and inconsistent with fast cooling synchrotron emission for the non-thermal part of the spectrum. 
%%Spectra which when fitted with Band function gave an $\alpha = -0.81$ were found to be best fitted with the slow cooling synchrotron function. At the same time the simulated synchrotron spectra were best fitted with the Band function with $\alpha = -0.8$, this enables to give a more physical meaning to the observed Band function spectral properties. 
%  

\cite{Burgess2014a} found that the blackbody temperature decreases as a broken power-law and  that the normalisation of the blackbody 
%%%%%
%%%%%
%%%%parameterised by ${\cal{R}} = (F_{\rm BB}/\sigma_T T^4)^{1/2}$, where $F_{\rm BB}$ is the observed blackbody flux and $\sigma_T$ is the Thompson scattering cross-section, 
%%%%%
%%%%%
 increases linearly with time for all bursts. These results are in agreement with the observations previously made by \cite{Ryde2005, Ryde&Pe'er2009, Axelsson2012}.  The spectral peak of the synchrotron emission, $E_{\rm sync}$, was %which is parameterised as 
%\begin{equation}
%E_{\rm sync} = \frac{3}{2} \frac{\hbar q \Gamma B}{m_e c (1+z)} \gamma_{\rm el}^2
%\end{equation}
%\noindent
%where  $B_{cr} = 4.41 \times 10^{13}$ Gauss is the quantum critical field, $m_e$ is the mass of the electron and $c$ is the speed of light. 
%The $E_{\rm sync}$ is
 found to decrease from hard to soft, as a broken power-law, for all the bursts.

\section{The Photosphere component and the Determination of the flow properties}
\label{outflow_cal}
%Photosphere + optically thin emission

{  The detection of a strong blackbody component suggests} that the flow is baryonic dominated as opposed to Poynting flux dominated, {  that is, the acceleration is predominantly done by the thermal pressure.} This is because for Poynting flux dominated outflows, {  using the standard assumption of constant reconnection rate \citep{Drenkhahn&Spruit2002}} %{  in the case of non-dissipative flow (no magnetic reconnection)}, 
the photospheric component is expected to {  be suppressed %(to a per-cent level contribution)  
\citep{Zhang&Pe'er2009,Hascoet2013} and %{  in the case of a dissipative flow,} 
have a peak energy larger than a {few} MeV \citep{Begue&Pe'er2015}. This is in contrast to the strong thermal components observed at $\sim 100$ keV in the
5 bursts studied here, which initially lie at around 40\%  (see %fig. \ref{fig:ratio} and 
\S \ref{r_0r_s}), apart from GRB110721A which only has a few per cent blackbody flux.  We point out that GRB110721A has been interpreted within a Poynting flux dominated model as well \citep{Gao&Zhang2015}, and it is possible that the magnetisation is non-negligible, though weaker, in the other bursts, where the thermal component is more pronounced.}

In this section, we consider the baryonic scenario in which the photosphere is formed at a radius above the saturation radius
following the standard (non-dissipative) fireball evolution \citep{Meszaros2006}.
%Most of the burst energy is then in the form of the kinetic energy of the outflow. 
%Depending on how far the photosphere is formed from the saturation radius, the thermal component will undergo adiabatic cooling due to continued expansion of the outflow. 
The flow is imagined to be advected through the photosphere, whose position is determined by the properties set by the central engine. The variability timescales observed for the bursts (on the order of the pulse width, $t_{\rm pulse}$) are  much longer than both the dynamical timescales and the typical widths of the time-bins used in the analysis. The former is the time the flow takes to reach the photosphere, $t_{\rm dyn} \equiv r_{\rm ph}/2\Gamma^2 c = 0.2 \; \rm ms$, where $r_{\rm ph} = 10^{\rm 12} \:\rm cm,\: \Gamma = 300$ is assumed and $c$ is the speed of light. Therefore, it is safe to assume that the central engine is  approximately steady and thereby  the flow is quasi-static over the duration of each time-bin.  

At some radius above the photosphere a fraction of the kinetic energy of the outflow is dissipated by some unspecified mechanism, accelerating the electrons to high energies. The likely distribution that is  expected from diffusive shocks is a Maxwellian - Boltzmann distribution with an extension of a power-law at high energies \citep{Baring1995, Spitkovsky2008}. 
Such an electron distribution is, indeed, consistent with the distribution required by the synchrotron fits done to the data \citep{Tavani1996, Burgess2014}. 

 %Depending on the electron distribution attained via shocks, the cooling of electrons can result in different shapes of the synchrotron spectrum. 

%
%The high energy electrons then cool off by radiating their energy via the optically thin non-thermal slow cooling synchrotron emission. 
%The spectral analysis done by \cite{Burgess2014a} shows that the non-thermal part of the spectrum is well fitted using a slow cooling synchrotron function. 
%This suggests that there is a balance between the synchrotron cooling of electrons and continuos heating or acceleration of the electrons for example in model with co -acceleration via first order and second order Fermi acceleration, see discussion for more details.  % and also in situations like envisaged in ICMART where there is a constant acceleration of electrons via second order turbulence.    

%\subsection{Outflow parameter calculations}

For each time bin, we estimate the outflow parameters $\Gamma$, $r_{\rm ph}$, $r_0$ and $r_{\rm s}$ by using the methodology described in \cite{Pe'er2007}: we use the blackbody's  temperature, $T$, its normalisation, which is parameterised by, 
\begin{equation}
{\cal{R}} =\left(\frac{F_{\rm BB}}{\sigma_{\rm SB} T^4}\right)^{1/2} =  \phi  \frac{(1+z)^2}{d_L} \frac{r_{\rm ph}}{\Gamma},
\label{R}
\end{equation}
where $\sigma_{\rm SB}$ is Stefan-Boltzmann constant and $\phi$ is a factor of the order of unity \citep{Pe'er2007}, $d_L $ is the luminosity distance
and $F$ is the total observed flux.  %, see also \cite{ Iyyani2013}).
We assume a redshift of  $z = 2$, the average value for GRBs \citep{Bagoly2006}, and assume a flat universe ($\Omega_{\Lambda} = 0.73, H_0 = 71$ km/s/Mpc).  The estimated outflow parameters change within a factor of a few for a different value of $z$, however, the time evolution of the behaviour of the parameters remain the same, see \cite{Iyyani2013}.  

{  We point out that we interpret the evolution of the flux and temperature as being due to central engine variations which causes evolution in the parameters. Alternative explanations involving high-latitude emission are discussed in \cite{Pe'er2008, Pe'er&Ryde2011}, and \cite{Lundman2013}.}

%\section{Implications }
\subsection{ Lorentz factor}
\label{Gamma}

\begin{figure*}
\begin{center}
\resizebox{84mm}{!}{\includegraphics{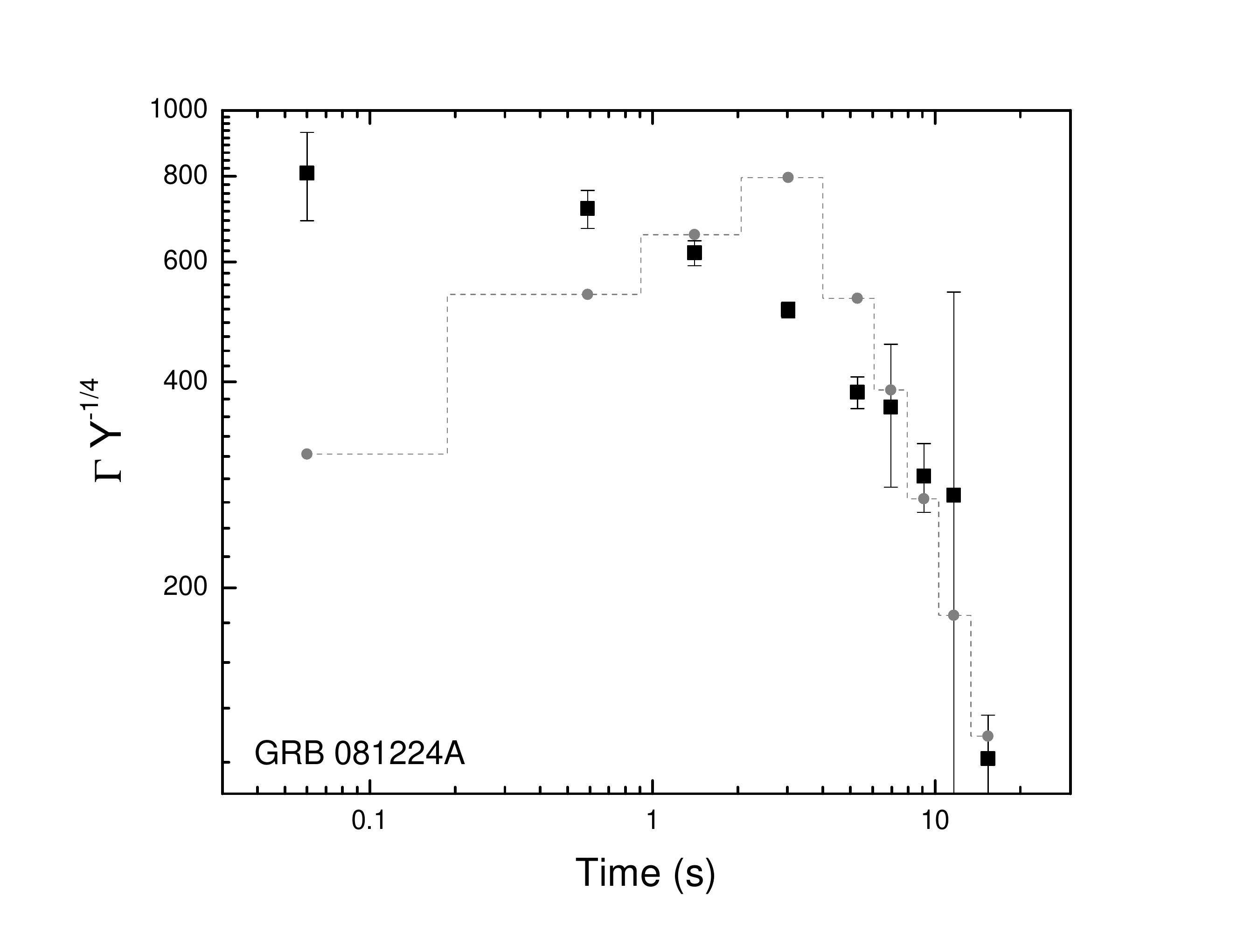}}
\resizebox{84mm}{!}{\includegraphics{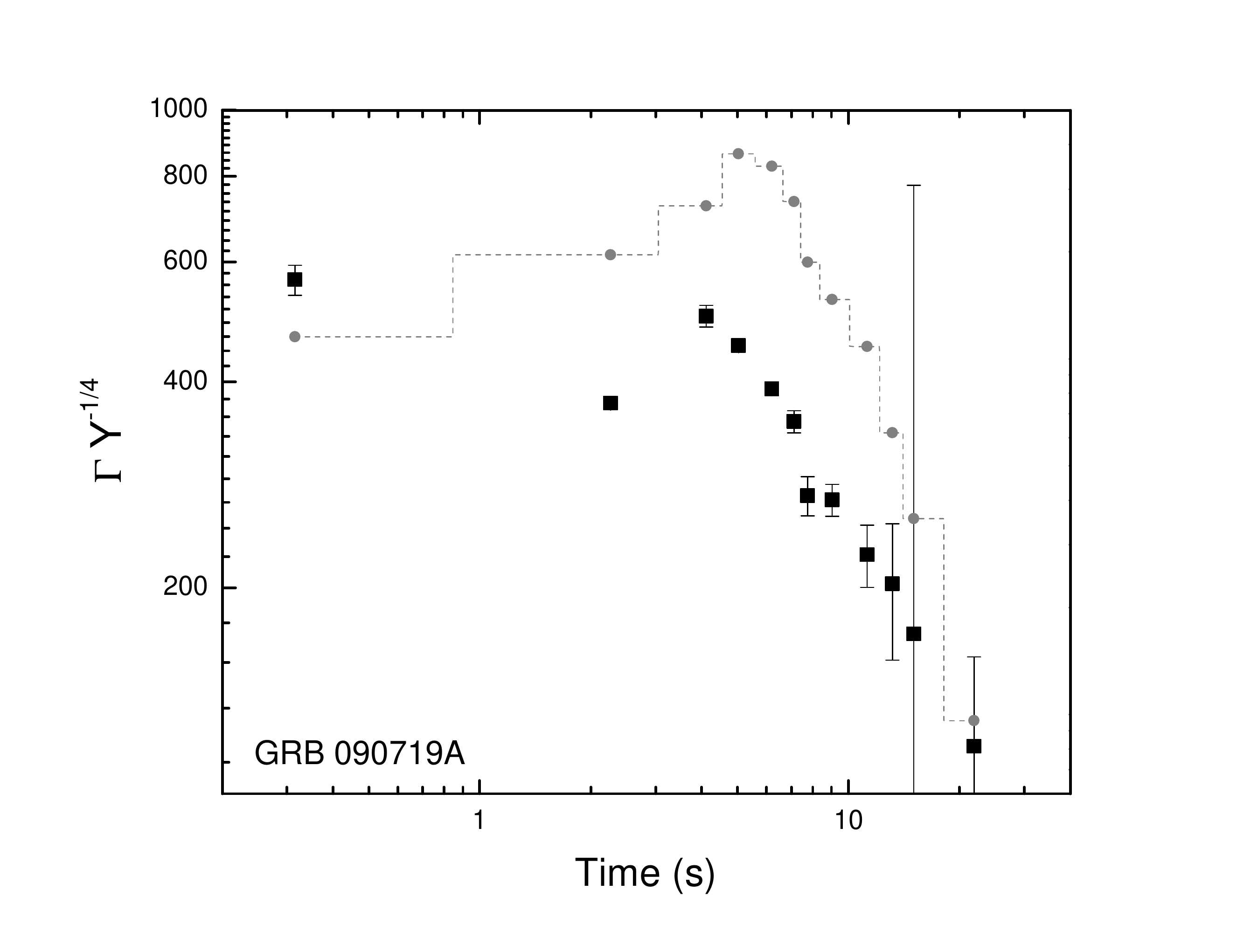}}
\resizebox{84mm}{!}{\includegraphics{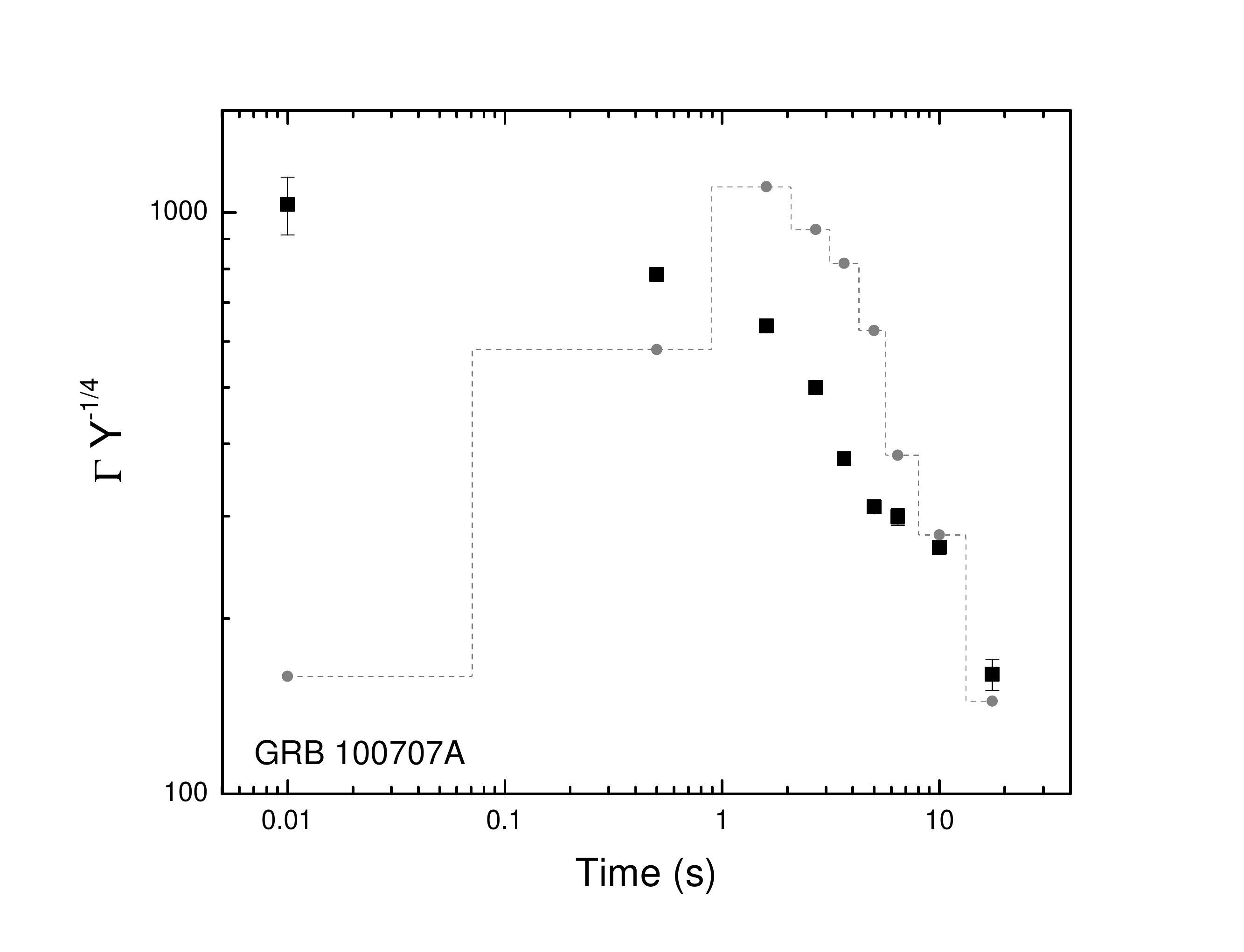}}
\resizebox{84mm}{!}{\includegraphics{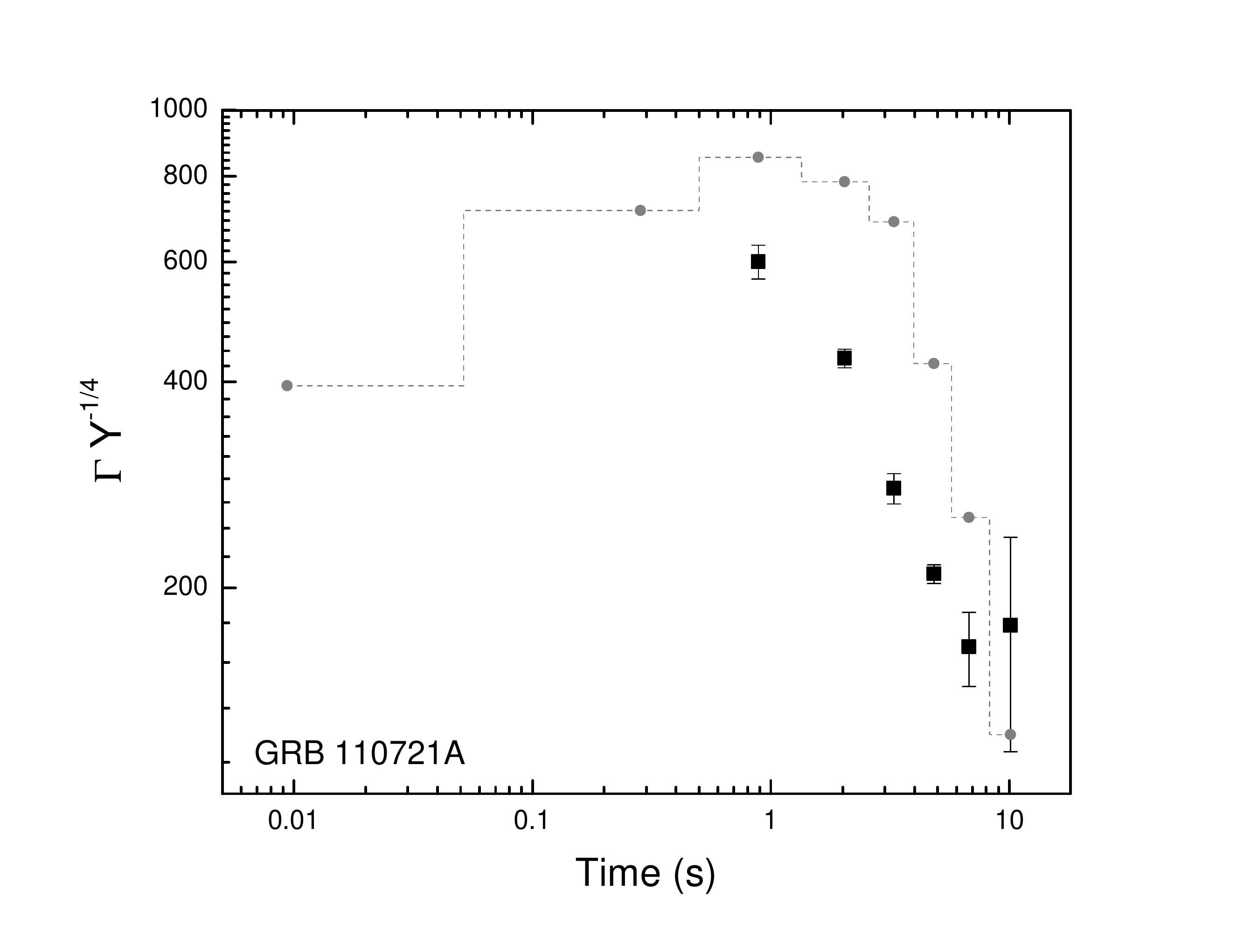}}
\resizebox{84mm}{!}{\includegraphics{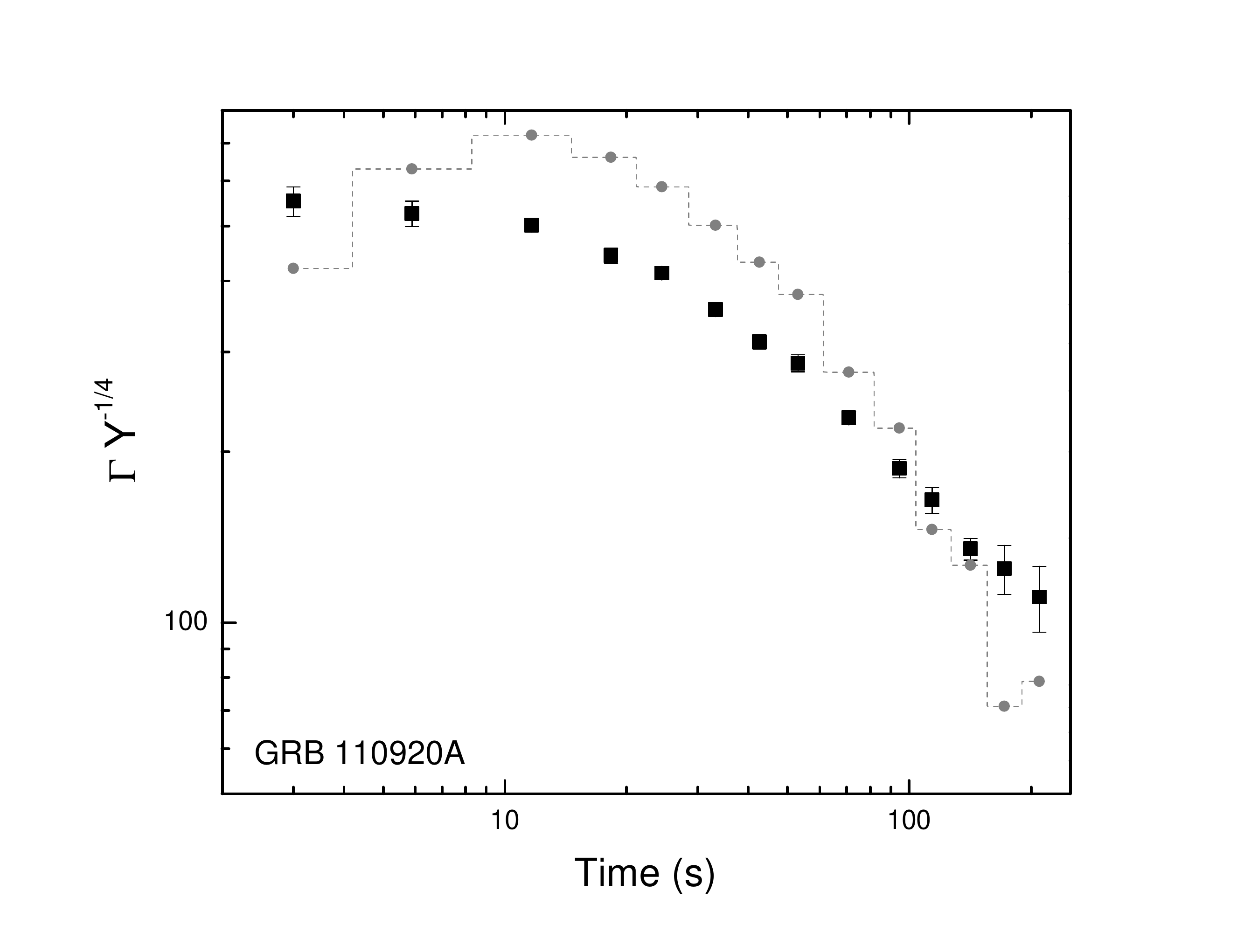}}
\caption{Evolution of Lorentz factor, $\Gamma \, Y^{-1/4}$, for five bursts, where $Y = {L_0}/{L_{\rm obs, \gamma}}$. The photon-flux light-curve (arbitrary units) is over-plotted in grey. Note that a blackbody component could not be detected during the first two time bins of GRB110721A.  }
\label{fig:gamma_phflux}
\end{center}
\end{figure*}

The Lorentz factor is found to decrease monotonously with time in {  the 5 bursts } which have a significant photospheric component, as shown in Figure \ref{fig:gamma_phflux}.  The Lorentz factor is derived from the observables by
\begin{equation}
\Gamma \propto (F/{\cal{R}})^{1/4} Y^{1/4}
\label{Gamma_eq}
\end{equation}
\noindent
where $Y$ % is the ratio of the total burst energy to the observed $\gamma -$ ray energy (
relates to the radiative efficiency of the burst which is given by 
\begin{equation}
Y = \frac{L_0}{L_{\rm obs, \gamma}}
\label{Y_rad}
\end{equation}
\noindent
where $L_0$ is the total kinetic luminosity and $L_{\rm obs, \gamma}$ is the observed $\gamma-$ray luminosity. 
Since during the rise phase of the pulse, both the total flux and  ${\cal{R}}$ increase with nearly a similar rate, 
%which represents the transverse size of the observed photosphere of the jet $(r_{\rm ph}/\Gamma)$, see equation \ref{R}, \citep{Ryde&Pe'er2009,Pe'er2007}, also increases with time. As a result, we find 
the Lorentz factor remains close to a constant or sometimes shows only a moderate decrease with time. However, during the decay phase  $\Gamma$ decreases much faster, since ${\cal{R}}$ continues to increase while the flux decreases. 

\begin{figure*}
\begin{center}
\resizebox{80mm}{!}{\includegraphics{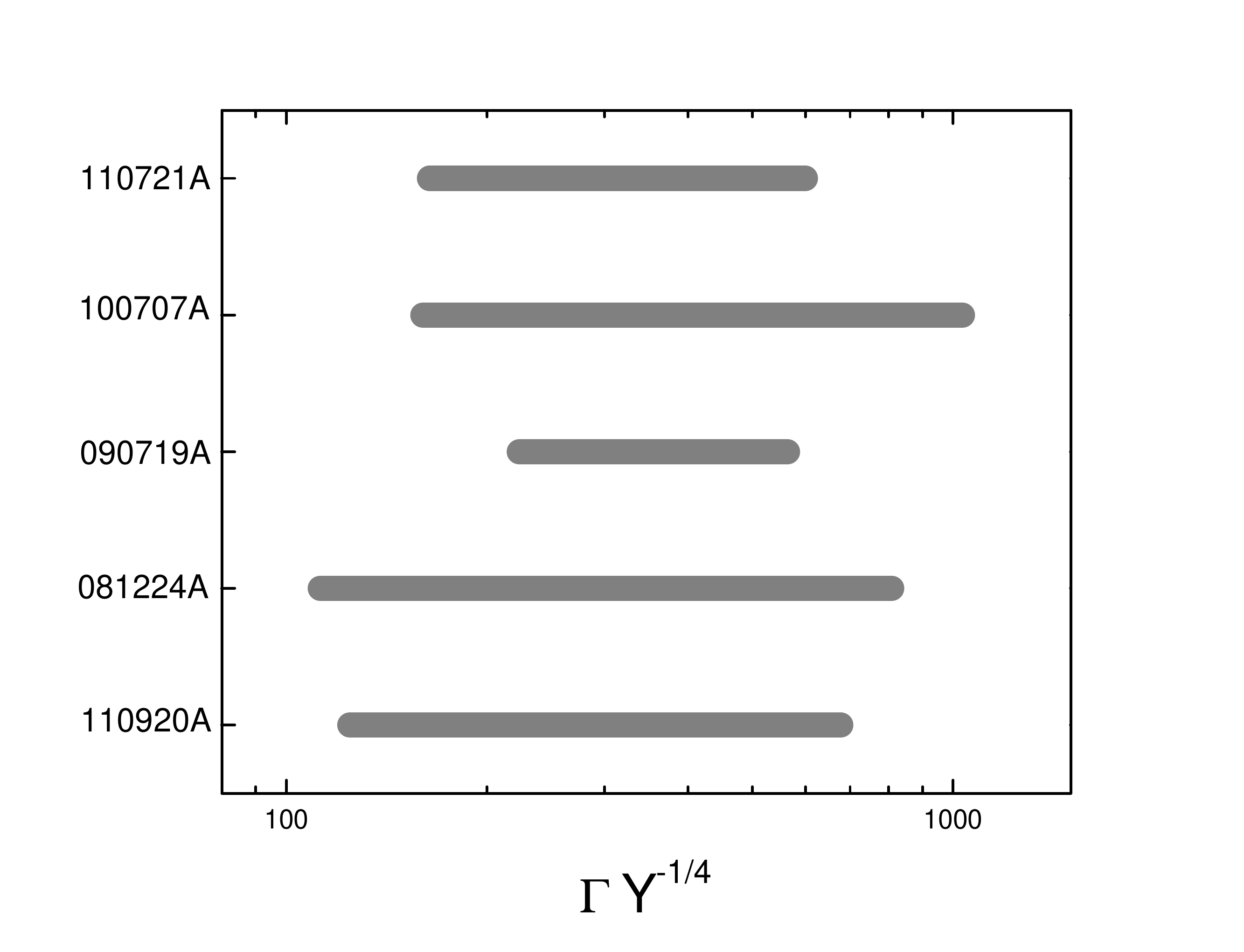}}
\resizebox{80mm}{!}{\includegraphics{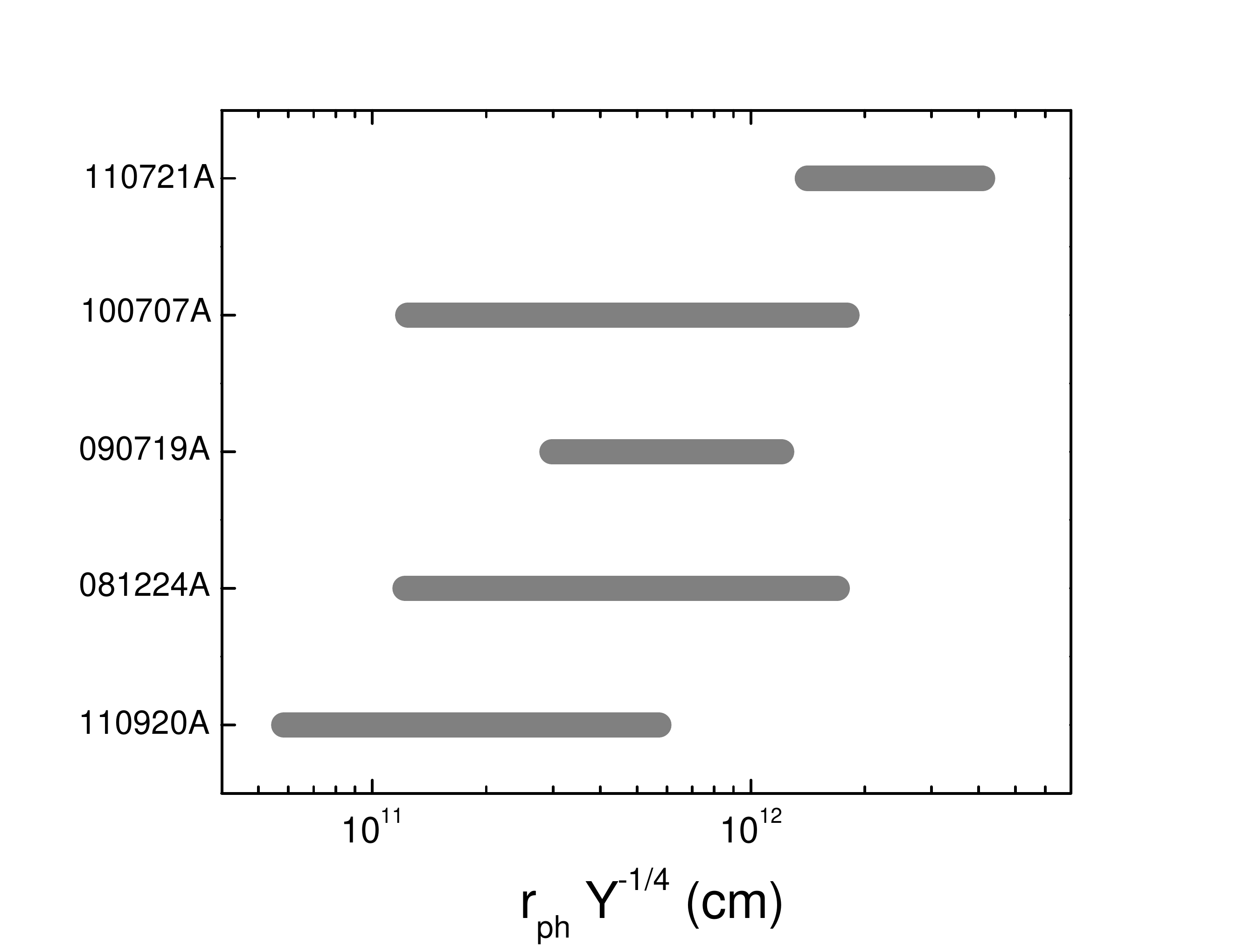}}
\resizebox{80mm}{!}{\includegraphics{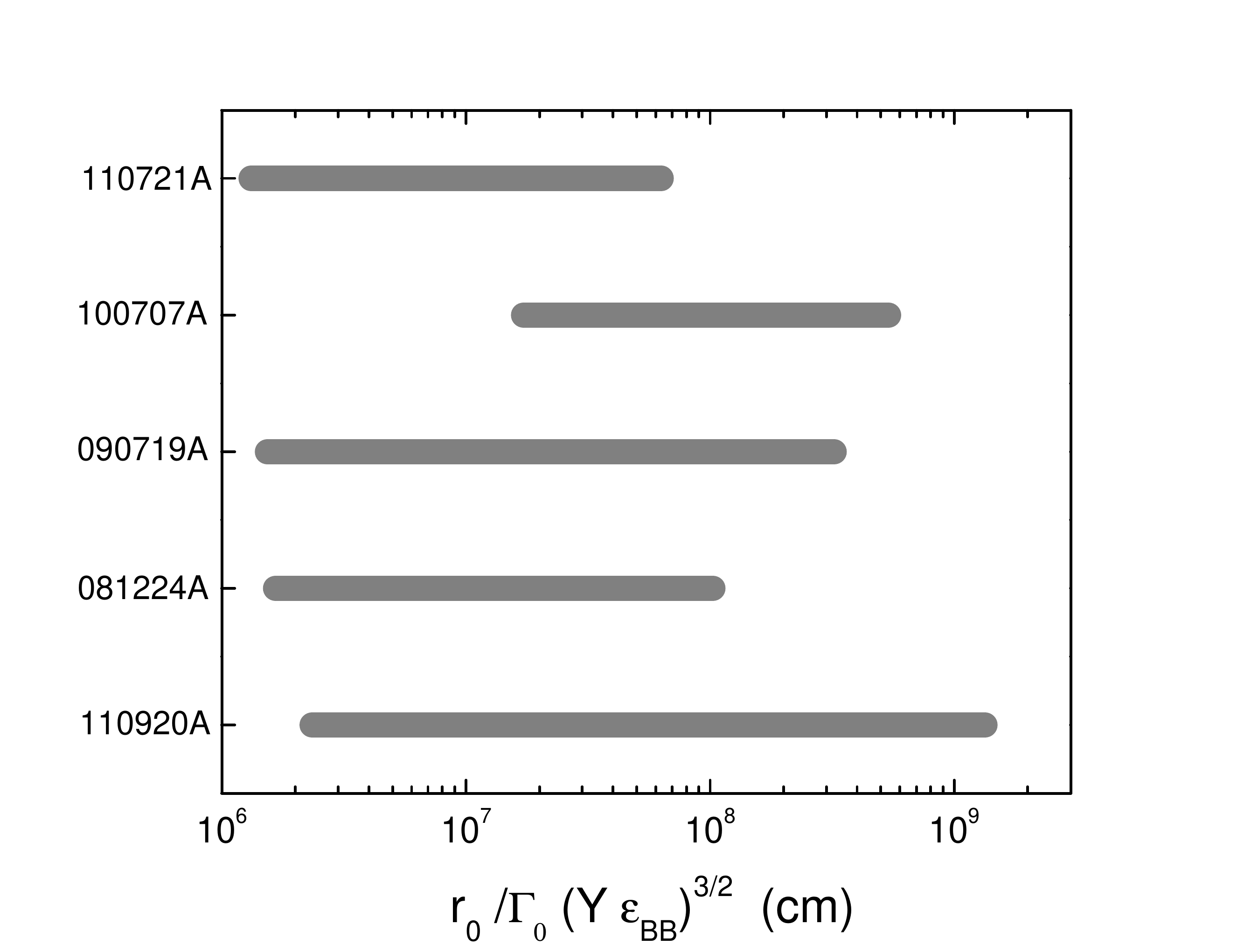}}
\resizebox{80mm}{!}{\includegraphics{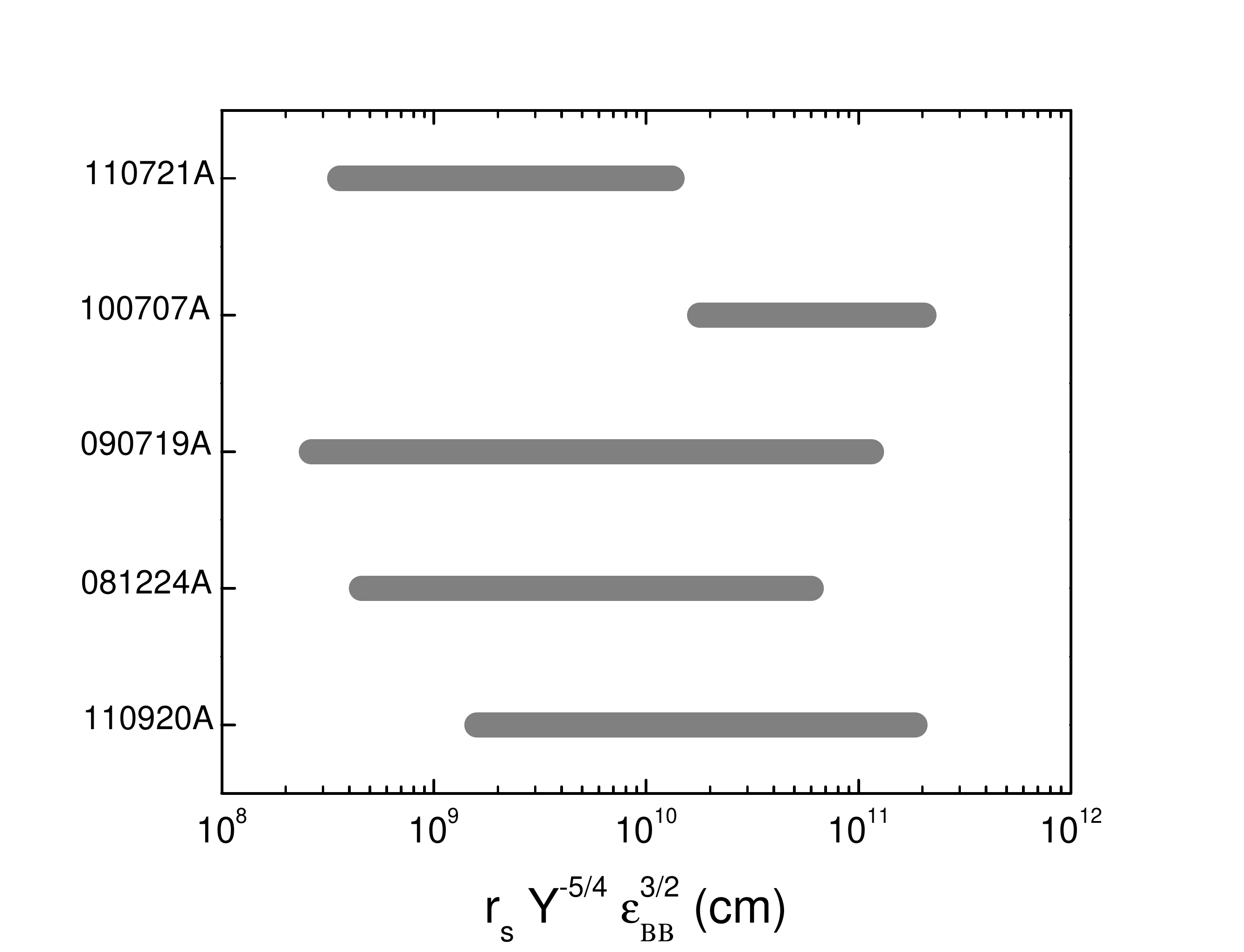}}
\caption{Ranges of the inferred outflow properties:  $\Gamma \, Y^{-1/4}$ (upper left-hand panel), $r_{\rm ph} Y^{-1/4}$ (upper right-hand panel), $r_{\rm 0}/\Gamma_{\rm 0} (Y \epsilon_{\rm BB})^{3/2}$ (lower left-hand panel) and $r_{\rm s} Y^{5/4}$ (lower right-hand panel) for an assumed redshift, $z = 2$.  Here $\Gamma_0$ is the Lorentz factor of the flow at $r_0$ and $\epsilon_{\rm BB}$ is the fraction of the burst energy that thermalises at the nozzle radius, $r_0$, of the jet (see \citet{Iyyani2013} for further details).}
\label{outflow_range}
\end{center}
\end{figure*}

 %In each burst, we find that during the rising phase of the pulse, $\Gamma$ is  at its maximal value and shows only a moderate decrease, while during the decay phase of the pulse, $\Gamma$ decreases more rapidly.  
 The upper left-hand panel in  Figure \ref{outflow_range} shows the range over which the Lorentz factor varies. All deduced values lie above 100 and only one burst (GRB100707A) has an initial value of $\Gamma$ greater than 1000. The average value of $\Gamma$ for the sample, considering their temporal evolution, is $\langle \Gamma \rangle = 377 \pm 205$.  
 
%{  Present very briefly, the other techniques that have been described in literature to determine $\Gamma$ which gives either the upper limit or lower limit. Compare the values of $\Gamma$ that has been deduced by other methods like break observed in afterglows (peak time of the optical afterglow); upper limit from the detection of LAT photons using the argument of $\gamma - \gamma$ pair production. Comparing with \cite{Racusin2011} results, we find, the estimates of $\Gamma$ we had made are well in agreement with the lower limits ($\sim 100$) and upper limits ($\sim 1000$) estimates.  }

%For the detection of the blackbody component, 

%\subsubsection{Behaviour of $\Gamma$ with respect to photon flux}

%{  $\Gamma$ evolves within the range of 1000 to 100; what does this signify; does it tell us anything about the central engine variation?}
% It is interesting to note that $\Gamma$ does not show an increase, reaching a peak and then decreasing, instead $\Gamma$ starts to decrease from an initially large value.  In other words, the rise time in $\Gamma$ is very small that it cannot be resolved to be deduced. 

%{  We can observe the rise phase in the pulse but not in $\Gamma$: what does that imply or mean?}

%\subsection{Implications}

\subsection{Photospheric radius}
\label{Rph}
The photosphere is the deepest region in the outflow from where photons can be observed. The derived values of the photospheric radius, $r_{\rm ph}$, is found to vary moderately with time (within a factor of ten), see Figure \ref{fig:radii}. The average value of $r_{\rm ph}$ for these bursts is $ \langle  r_{\rm ph} \rangle= 10^{11.8 \pm 0.4}$ cm, and the deduced ranges of values for different bursts are shown in the upper right-hand panel in  Figure \ref{outflow_range}. 
 
The dependance of  $r_{\rm ph}$ on the observables is given by 
\begin{equation}
r_{\rm ph} \propto L_0/\Gamma^3 \propto \dot M/ \Gamma^2
\end{equation} 
\noindent
where $\dot M$ is the baryon load and is given by 
%\begin{equation}
$\dot M = {L_0}/{\Gamma c^2}$.
%\label{Mdot}
%\end{equation}
%\noindent
The evolution of the photospheric radius thus depends on a combination of the evolution in baryon load and Lorentz factor: the increase in baryon load increases the particle density, whereas the Lorentz factor effects the optical depth through
$\tau = (n_e \sigma_T R)/2\Gamma^2$,  where $n_e$ is the electron number density {  at $R$,} {  which is the 
distance of the emission site from the central engine, and  $\sigma_T$ is the Thompson cross-section}.  
In particular,  the evolution of $r_{\rm ph}$ is very sensitive to the changes in $\Gamma$. 

During the rise phase of the pulse, the variation in $r_{\rm ph}$ is mainly given by the variation in the luminosity of the burst, since $\Gamma \sim$ constant (\S \ref{Gamma}). This corresponds to an increase of the baryon load 
which causes the opacity, due to the electrons associated with the baryons, to increase, thereby increasing the photosphere radius. {  However, since the luminosity of the burst shows a corresponding increase, the energy per baryon (i.e $\Gamma$) remains nearly constant during the rise phase of the GRB pulse.}  

On the other hand, during the decay phase of the pulse, both $\Gamma$ and $L_0$ decrease, whereas $r_{\rm ph}$ is found to be nearly steady or moderately increasing. In order to keep $r_{\rm ph}$ steady, it requires that $L_0 \propto \Gamma^3$ (or  $\dot M \propto \Gamma^2$). %{  On the other hand, to have an increase in $r_{\rm ph}$ during the decay phase of the pulse, there should be either an increase in the baryon load 
%( see  eq. \ref{Mdot}) which can be obtained when $\Gamma$ decreases faster than $L_0$; 
%or in a scenario where $\dot M$ remains a constant or decreases with time, 
% In such a case, it is still possible for $r_{\rm ph}$ to increase with time, when 
%due to the decrease in $\Gamma$, there should be an increase in the  CAN THIS BE SHORTENED?}

%However, during the decay phase, both $\Gamma$ and  $L_0$ decrease, whereas $r_{\rm ph}$ is found to be nearly steady or moderately increasing. This suggests that  $L_0 \propto \Gamma^3$. On the other hand, if we assume the baryon load ($\dot M $) remains a constant throughout, then the increase observed in $r_{\rm ph}$ is mainly due to the decrease in $\Gamma$. This is because, as $\Gamma$ decreases
%the optical depth, $\tau = (n_{\rm e} \sigma_{\rm T} R)/2\Gamma^2$, increases with time and this in turn results in a farther position of the photosphere with time. 

%As $\Gamma$ decreases, the relative distance travelled by the electrons with respect to the photons along the radial motion of the photon, decreases. As a result, the size of the region where the photons and electrons interact increases, thereby resulting in larger optical depth. In other words, the effective electron number density increases. 
%where $n_e$ is the electron number density. 
%Thus, the increase in optical depth results in a farther position of the photosphere with time. 

%{   What does the evolution of the $r_{ph}$ signify: almost steady photospheric radius  and increasing photospheric radius ?} 

\subsection{Nozzle and saturation radii}
\label{r_0r_s}
\begin{figure*}
\begin{center}
\resizebox{84mm}{!}{\includegraphics{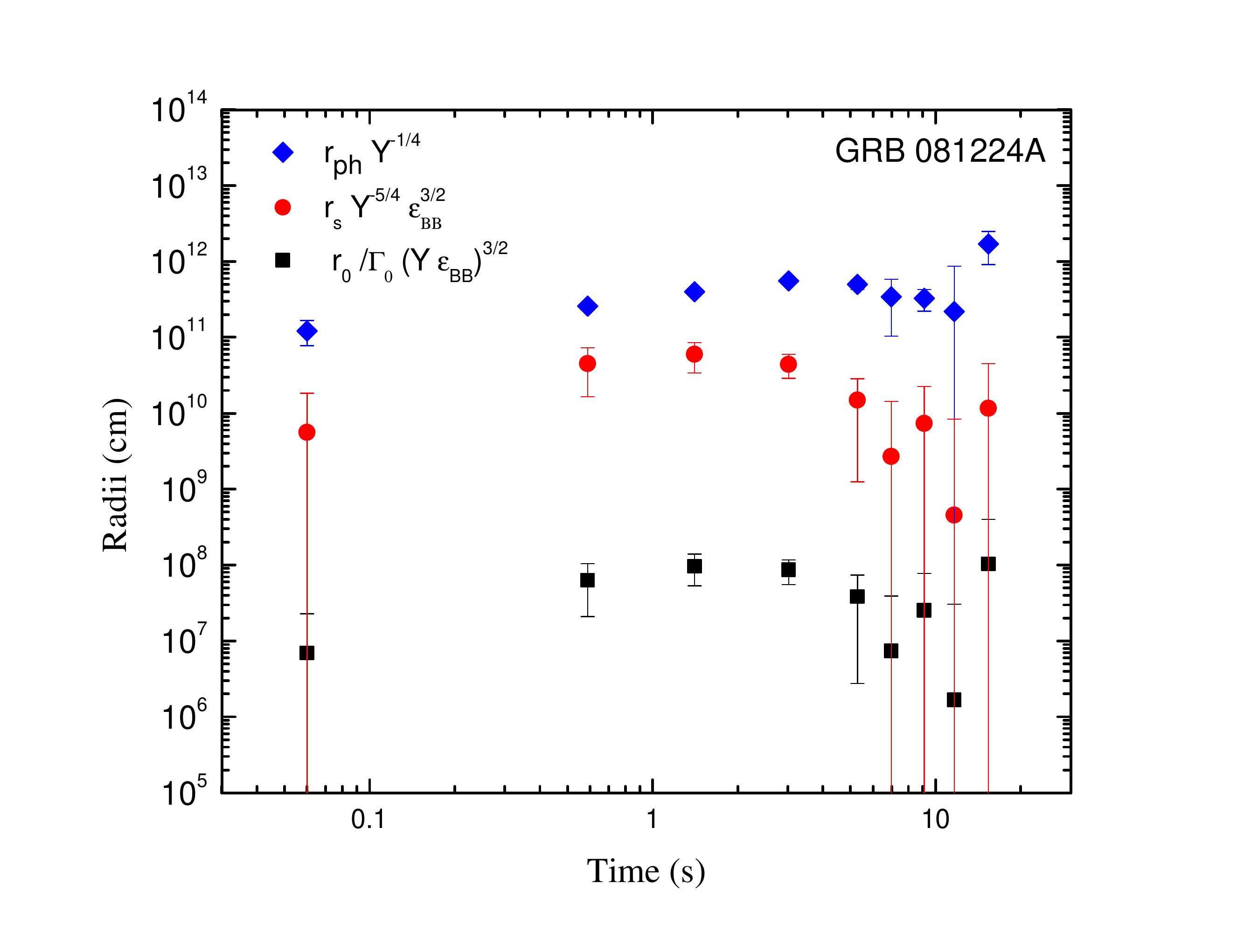}}
\resizebox{84mm}{!}{\includegraphics{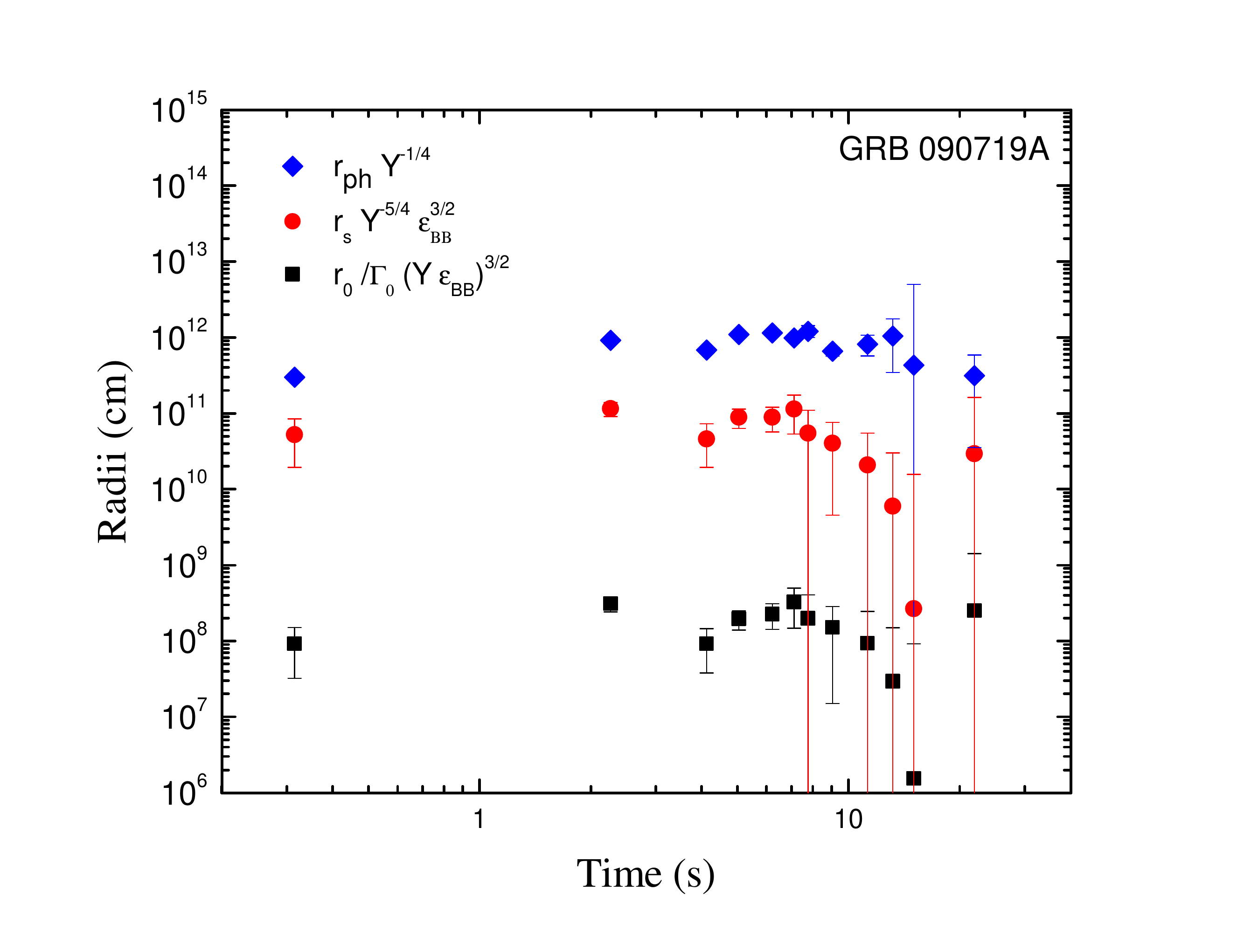}}
\resizebox{84mm}{!}{\includegraphics{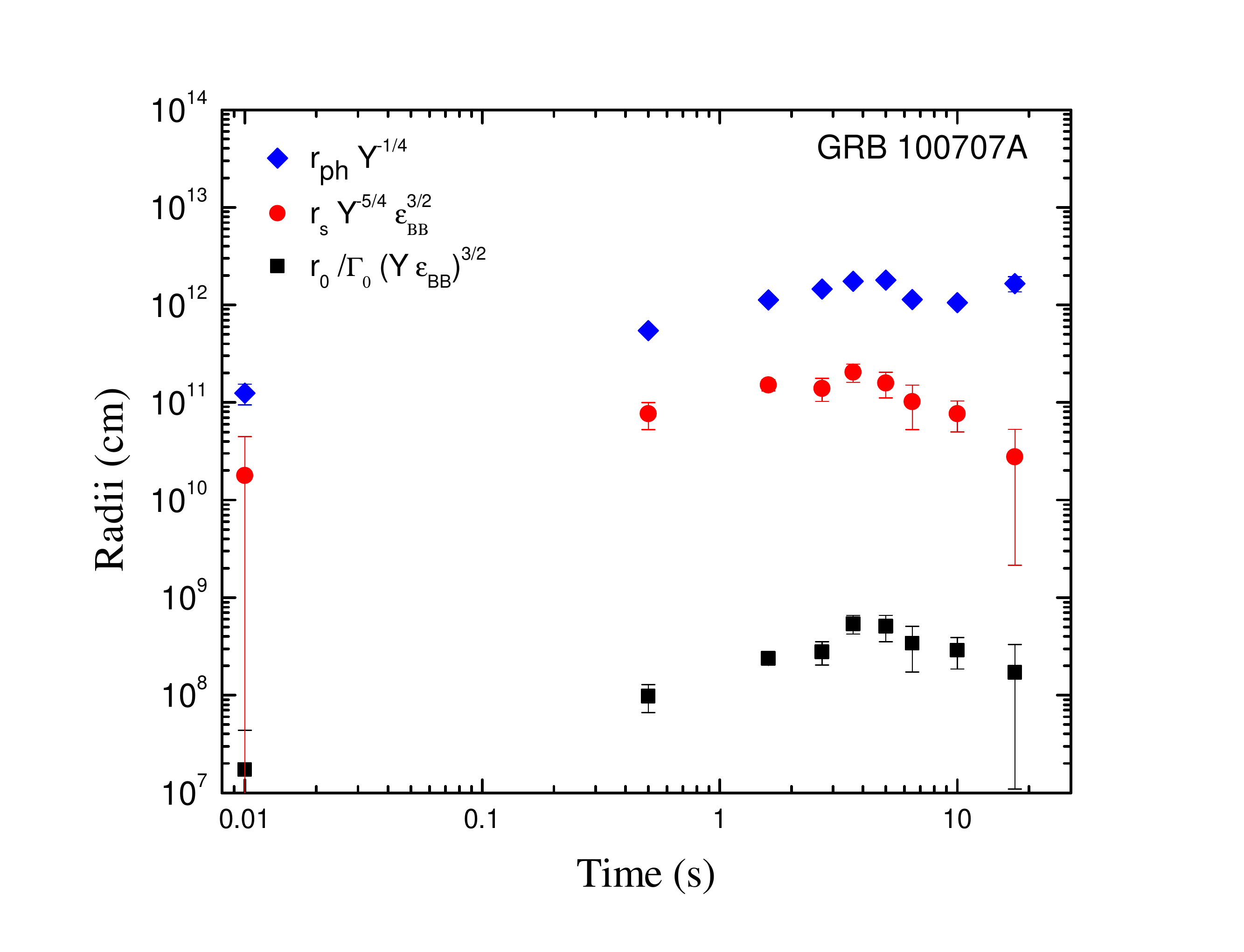}}
\resizebox{84mm}{!}{\includegraphics{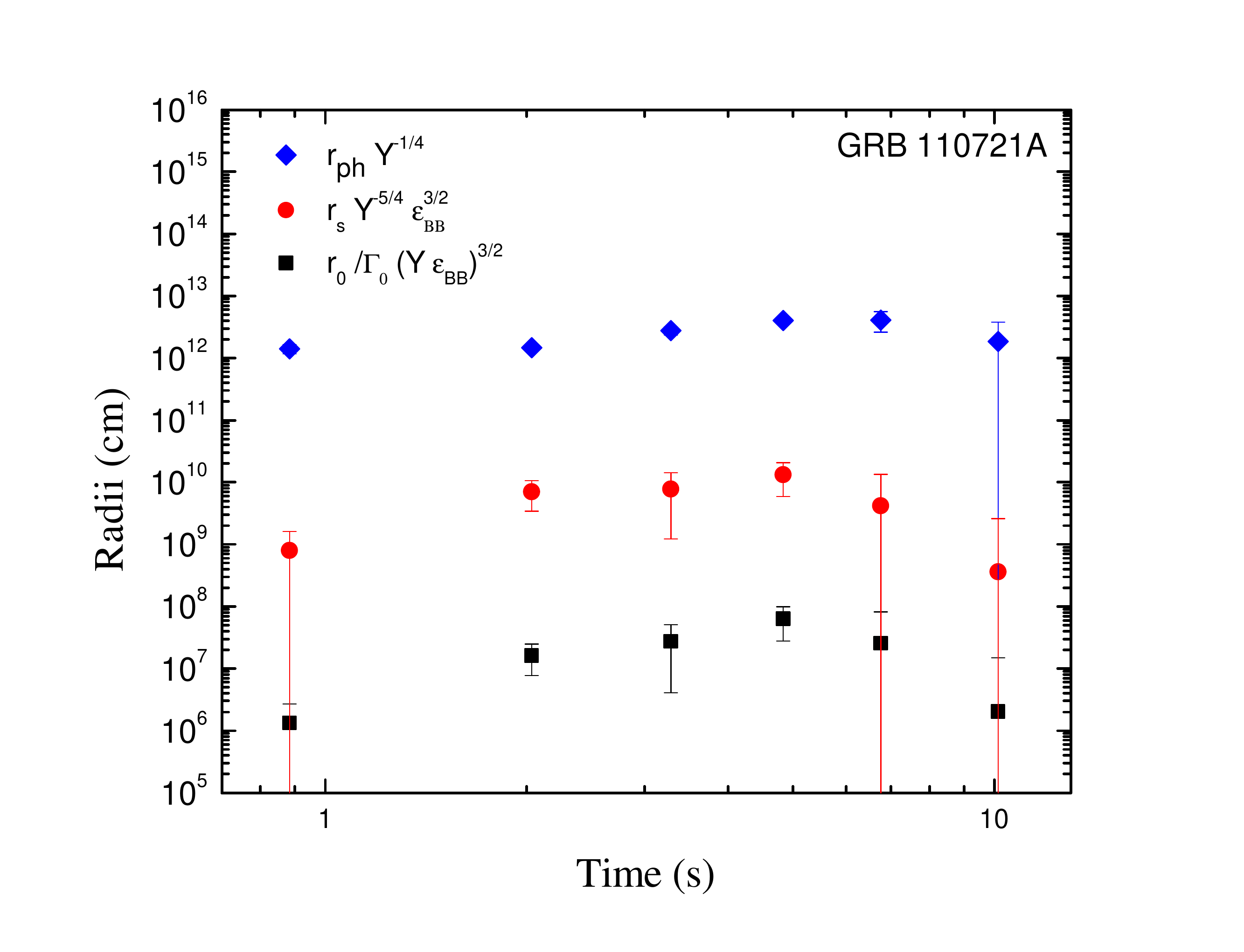}}
\resizebox{84mm}{!}{\includegraphics{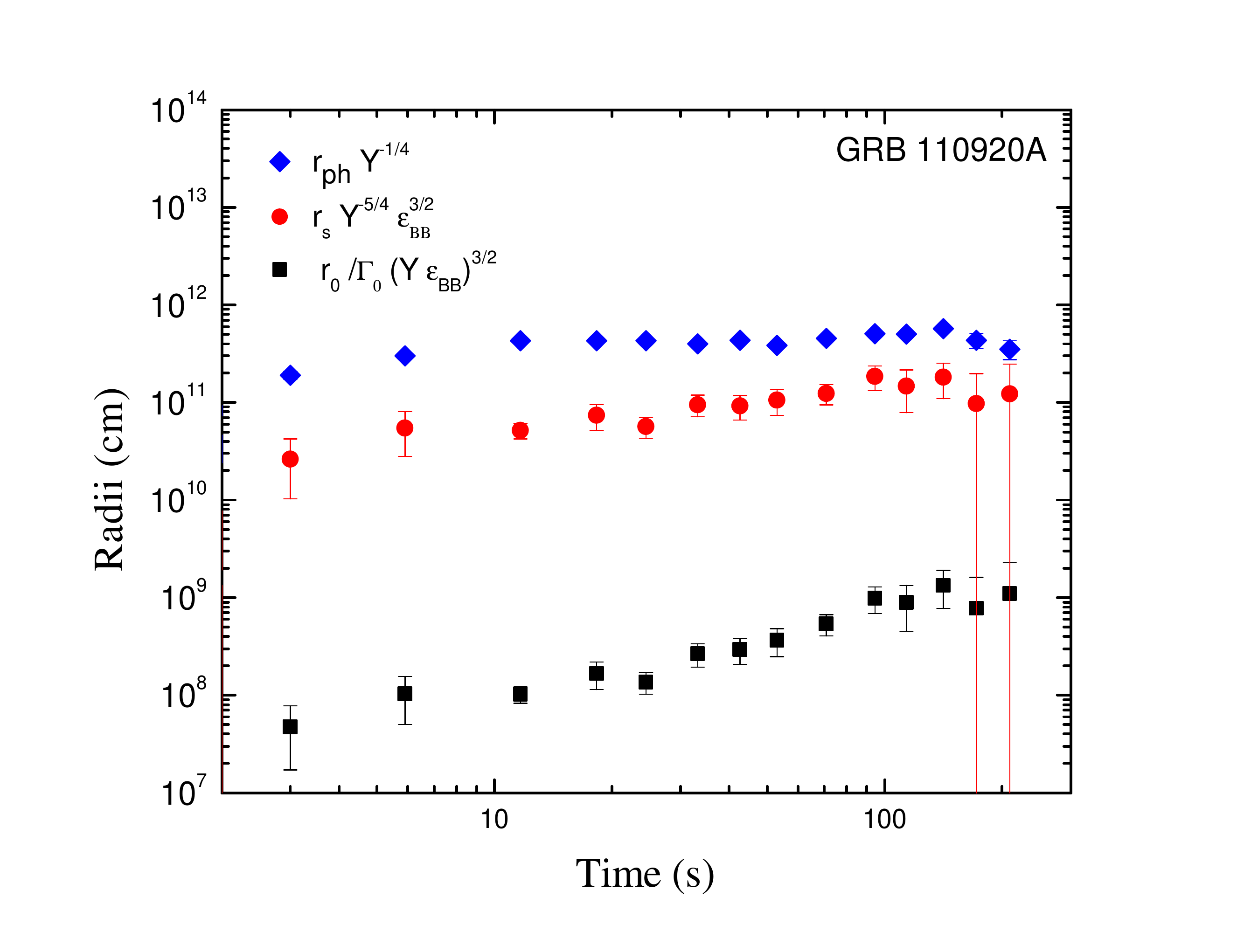}}
\caption{The evolution of the photospheric radius, $r_{\rm ph} Y^{-1/4}$ (blue diamonds), saturation radius, $r_{\rm s} Y^{5/4} \epsilon_{\rm BB}^{3/2}$ (red dots) and nozzle radius, $r_{\rm 0}/\Gamma_{\rm 0} (Y \epsilon_{\rm BB})^{3/2}$ (black squares). For details, see Figure 2.}
\label{fig:radii}
\end{center}
\end{figure*}

\begin{figure}
\begin{center}
\resizebox{84mm}{!}{\includegraphics{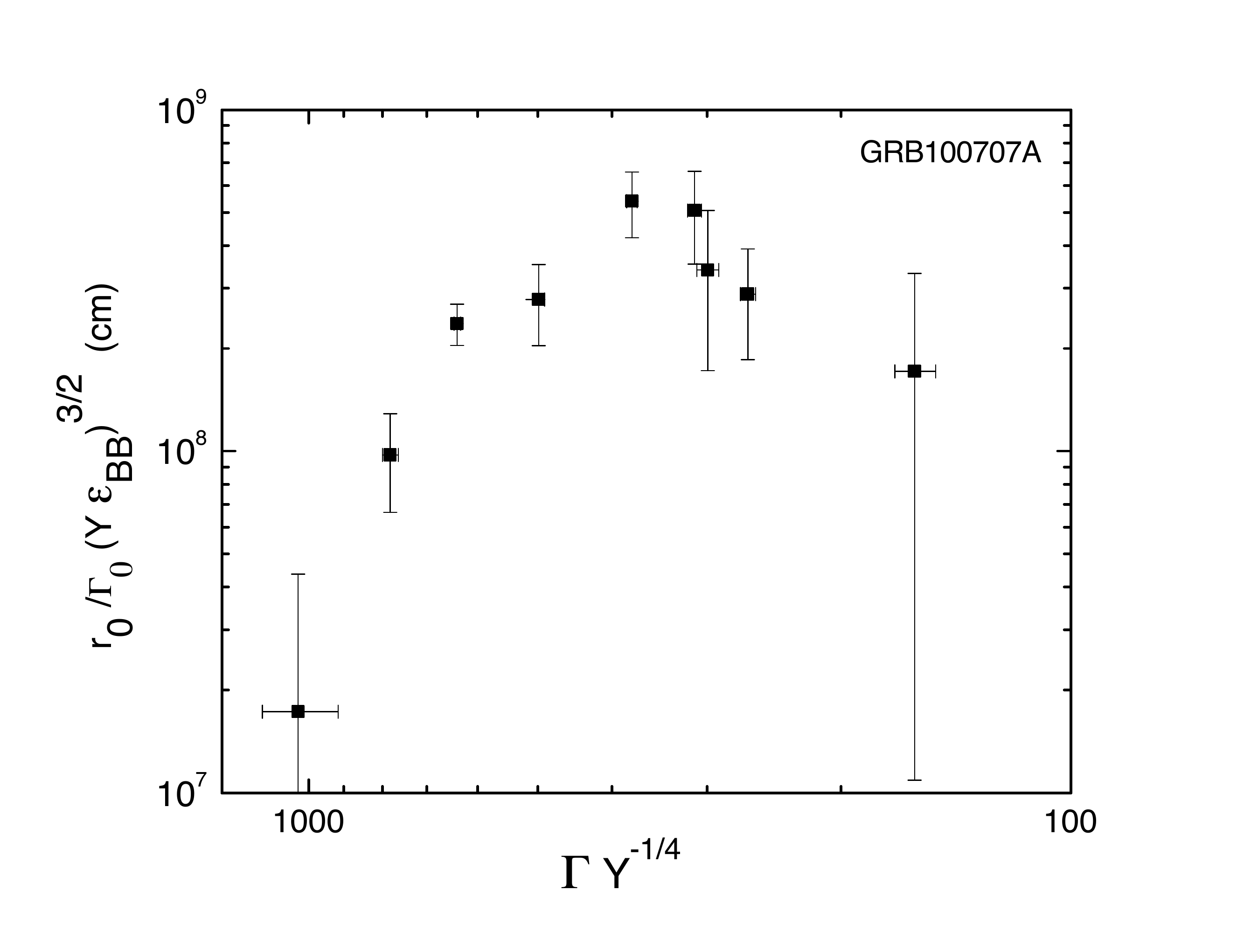}}
\caption{The correlation between $r_0$ and $\Gamma$ for GRB100707A. The break in the correlation is mainly due to the break in the evolution of $r_0$ occurring at  $\sim 5$ s (see Fig. \ref{fig:radii}).}
\label{gamma_r0}
\end{center}
\end{figure}

The nozzle radius of the jet, $r_0$, signifies the radius from where the jet starts to accelerate\footnote{{  Note that the radius $r_0$ does not necessarily need to correspond to the radius of the central engine, and instead can be related to the dissipation pattern of the flow within the star \citep{Thompson2007, Iyyani2013, Pe'er2015}.}}.  The bursts show a common behaviour according to which $r_0$ initially increasing by a factor of approximately ten (Fig. \ref{fig:radii}). After a few seconds, $r_0$ decreases {again reaching its original value}.  One exception is, however, GRB110920A for which the $r_0$--break occurs after 100 seconds.  We note that maximal value of $r_0$ is not attained in coincidence with the peak of  the light curve, but is attained during the decay phase (with the exception of GRB081224A where the $r_0$-peak occurs just before the light-curve peak.)
The  lower left-hand panel in Figure \ref{outflow_range} shows that the observed value of $r_0$ varies within the range of $10^6$ to a few $\times 10^9$ cm, with an average value, $ r_{\rm 0, av} = 10^{7.9 \pm 0.8}$ cm.  These values are well within the expected range between the black hole event horizon radius  [$10^{6 -7} \: \rm cm$ for a black hole mass of $5 -10 \: M_{\sun}$ \citep{Paczynski1998}] and the size of the core of an expected progenitor Wolf-Rayet  star \citep{Woosley&Weaver1995}. This suggests that $r_0$ is related to the interaction between the jet and the progenitor star \citep{Thompson2007, Iyyani2013}{ , see further discussion in section \ref{r_0disc}}. Figure \ref{gamma_r0} shows the correlation between $r_0$  and $\Gamma$. The break in the correlation is mainly due to the break in the evolution of $r_0$ occurring during the pulse decay phase.

The pulse-like temporal behaviour in $r_0$ gets reflected in the behaviour of $r_{\rm s}$, see Figure \ref{fig:radii}. For the bursts in the sample, $r_{\rm s}$ has an average value $ \langle r_s \rangle = 10^{10.5 \pm 0.8}$ cm. 
The ratio $r_{\rm s}/r_{\rm ph}$ also varies with time. This ratio is given by $(F_{\rm BB} / F_{\rm kin})^{3/2}$, where $F_{\rm BB}$ is the blackbody flux and $F_{\rm kin}$ is the kinetic energy flux at the photosphere (eq. 1 in \citet{Iyyani2013}). 
Figure \ref{fig:ratio} shows the ratio of the measured blackbody flux and the total flux (blackbody + synchrotron), which generally  is a good approximation for $F_{\rm BB} / F_{\rm kin}$, {  provided $Y$ is close to unity. Since $Y$ is unknown, these ratios are upper limits.}
The ratio {  initially lies around 40\%  (except for GRB110721A, see further discussion in \S \ref{sec:Poynting})} and typically decreases towards the end of the burst (see also \cite{Ryde&Pe'er2009}).  For GRB110920A,  which has a significantly different pulse length, the ratio is shown in Figure 7 in \cite{Iyyani2015} and has a similar behaviour, with the ratio lying between 10 and 40$\%$\footnote{Iyyani et al. (2015) argues for a different physical interpretation for this bursts, including thermal Comptonisation due to localised dissipation below the photosphere, see also \S \ref{sec:alternative}.}. {  The blackbody components are thus very strong in four of the bursts and are comparable, to within a factor of two, to the archetypal baryonic photosphere burst GRB090902B which has an average ratio of 70\% \citep{Ryde2010, Pe'er2012}.}

\section{Synchrotron component}
\label{sync}
In this section, we study the variation observed in the properties of the non-thermal component, that is, the synchrotron component.

\subsection{Observed Spectral Behaviour %Evolution of the synchrotron peak, $E_c$
}
\cite{Burgess2014a} found that to produce acceptable spectral fits,  a blackbody was required in addition to the synchrotron component
in 5 of the bursts  (see \S \ref{Sample}). In these cases, the C-stat improves by values $> 10$ in comparison to the synchrotron only fits, see Table 2 and 3 in \cite{Burgess2014a}. The presence of a blackbody in these bursts were found to be statistically significant, verified through simulations.  The observed synchrotron peak, $E_{\rm sync}$, and the blackbody temperature, $T$, evolve as a broken power-law functions. 
%{  The E peaks broken power law}
In bursts GRB081224A, GRB090719A and GRB100707A the two breaks are consistent, while in GRB110721A and GRB110920A the breaks occur at different times.  %{  Why is this difference? Note: GRB110721A behaviour of k$T$ is consistent with what was observed in the analysis of Axelsson et al. 2012 however not Epeak. What could be the implication of this? What does the simultaneity of the breaks imply? However, in a few of those cases, we observe a difference in the slopes of evolution before and after break, what does that imply?}  

On the other hand, there are three bursts (GRB081110A, GRB110407A and GRB090809A) that are consistent with a synchrotron component alone, where the improvement in C-stat for the addition of a blackbody component is less than $10$. In these cases, the presence of a blackbody component could not be statistically validated by simulations. These bursts are also very bright and are not different from the other bursts  in terms of their properties, such as peak energy and fluence. The observed synchrotron peak in these cases also evolve as a broken power-law. %Check!}

%For all the fits, synchrotron emission from a fast-cooled electron distribution is clearly rejected by the data due to the broad curvature of the spectrum imposed by such a distribution. However, synchrotron emission from an uncooled electron distribution  

%{  In all bursts the synchrotron component is given by slow cooling.É 5 with BBÉ }

%Change in Cstat for the BB addition. And in the cases for synchrotron only.
%In three cases synchrotron only. See figure of the change in Cstat.

%The synchrotron peak energy is given by  $E_{\rm sync} = E_* \gamma_m^2$ where $E_* = (3 \Gamma B m_e c^2 )/(2 B_{cr} )$, and $B_{cr} = 4.41 \times 10^{13}$ cm.  In the fitting routine, we determined $E_*$ and $\gamma_m$ was fixed at $30$.  We find that in all the bursts in the sample $E_{\rm sync}$ evolves as a broken power law. 

%{  What is the difference between the synch only cases and the B synch cases?}

Before discussing the constraints that these fits give, we review the basic points of synchrotron emission.

%\subsubsection{Synchrotron peak $E_{*}$ and $\Gamma$} 
%Expected  Correlation of the model.

 %which we find has a parameter space of the order of $10^{-1} - 10^{4} \: \rm G$ \cite{Beniamini2013;Uhm&Zhang2013}. 
%We also find that $B$ decreases with time by an order of magnitude, following the relation of the synchrotron energy peak, $E_{*}$. 

\subsection{Synchrotron emission}

The dissipation of the kinetic energy of the outflow at a certain radius, $r_d$, causes the electrons to be accelerated to some characteristic Lorentz factor, $\gamma_{\rm el}$. The observed peak energy of synchrotron emission from these electrons is given by  (see, e.g., \cite{Rybicki&Lightman1986}) 

\begin{equation}
E_{\rm sync} = \frac{3}{2} \hbar \frac{qB}{m_e c} \gamma_{\rm el}^2 \frac{\Gamma}{(1+z)} %=  2.8 \times 10^{-20} B \gamma_{el}^2 \frac{\Gamma}{(1+z)} \rm erg     
\label{E_c}
\end{equation}
\noindent
where $B $ is the magnetic field intensity in the comoving frame, $m_e$ and $q$ are the mass and charge of an electron, respectively, and $\hbar$ is the reduced Planck constant. The observed synchrotron flux is given by 
\begin{equation}
F_{\rm sync} = \frac{\sigma_T c \Gamma^2 \gamma_{\rm el}^2  B^2 N_e}{24 \pi^2 d_L^2}
\end{equation}
\noindent
where $N_e$ is the number of radiating electrons. 

%%{{  what is the value of $\gamma_min$?  Bosnjak et al.}}

As the electrons emit radiation, they cool, and the radiative cooling time is given by \citep{Sari1996}

\begin{equation}
t_{\rm cool}  = \frac{6 \pi m_e c}{\sigma_T B^2 \Gamma \gamma_{el} (1+{\cal{Y}})} \\
\label{tcool}
\end{equation}
\noindent
where %$U_{\rm B} = \rm B^2 / 8\pi$ is the energy density in the magnetic field, 
${\cal{Y}}$ is the Compton ${\cal{Y}}$-parameter and the factor $(1+{\cal{Y}})$ takes into account the cooling due to Compton scattering. 
The cooling time in equation (\ref{tcool}) can be compared to the dynamical time, 
\begin{equation}
t_{\rm dyn} \simeq \frac{R}{2 \Gamma^2 c}.
\label{tdyn}
\end{equation}
\noindent
Here, we assume that the conditions change on a timescale comparable to the dynamical timescale of the dissipation event. 
If $t_{\rm cool} <  t_{\rm dyn}$ the electrons radiate efficiently and loose all their energy within the dynamical time (fast cooling) and if $t_{\rm cool} >  t_{\rm dyn}$ the electrons do not efficiently radiate and thereby do not loose their energy during the dynamical time (slow cooling). {  The radiative efficiencies of the bursts in our sample are unknown. However, in recent years there is accumulating evidence, based on afterglow measurements, indicating that the efficiency of the prompt phase might be very high (\citet{Cenko2010,Racusin2011,Wygoda2015}, however see \cite{Santana2014} and \cite{Wang2015}).}

\begin{figure}
\begin{center}
\resizebox{84mm}{!}{\includegraphics{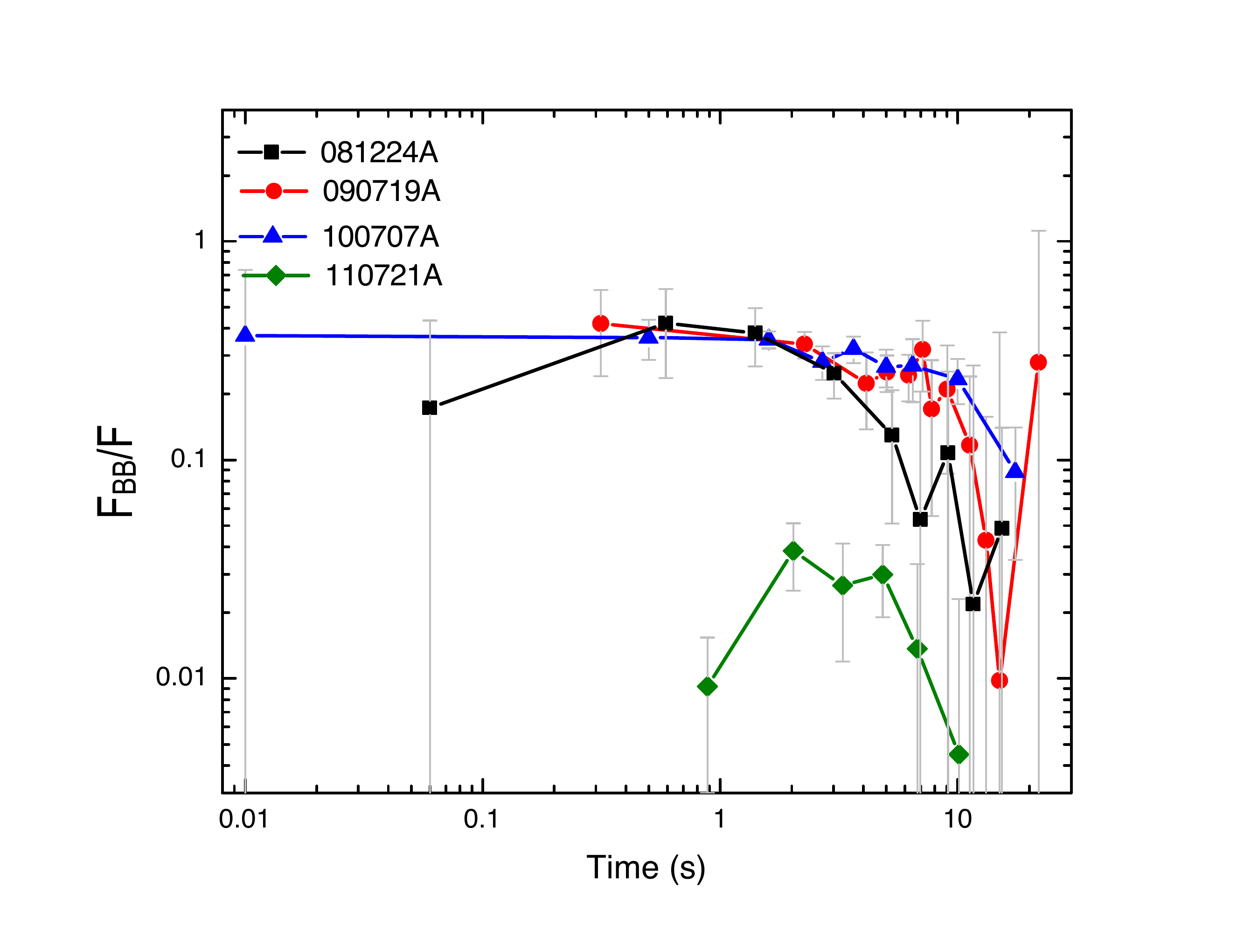}}
\caption{The ratio between the blackbody flux and the total flux (blackbody + synchrotron). A corresponding plot for GRB110920A is given in Figure 7 in Iyyani et al. (2015).}
\label{fig:ratio}
\end{center}
\end{figure}

\begin{figure}
\begin{center}
\resizebox{84mm}{!}{\includegraphics{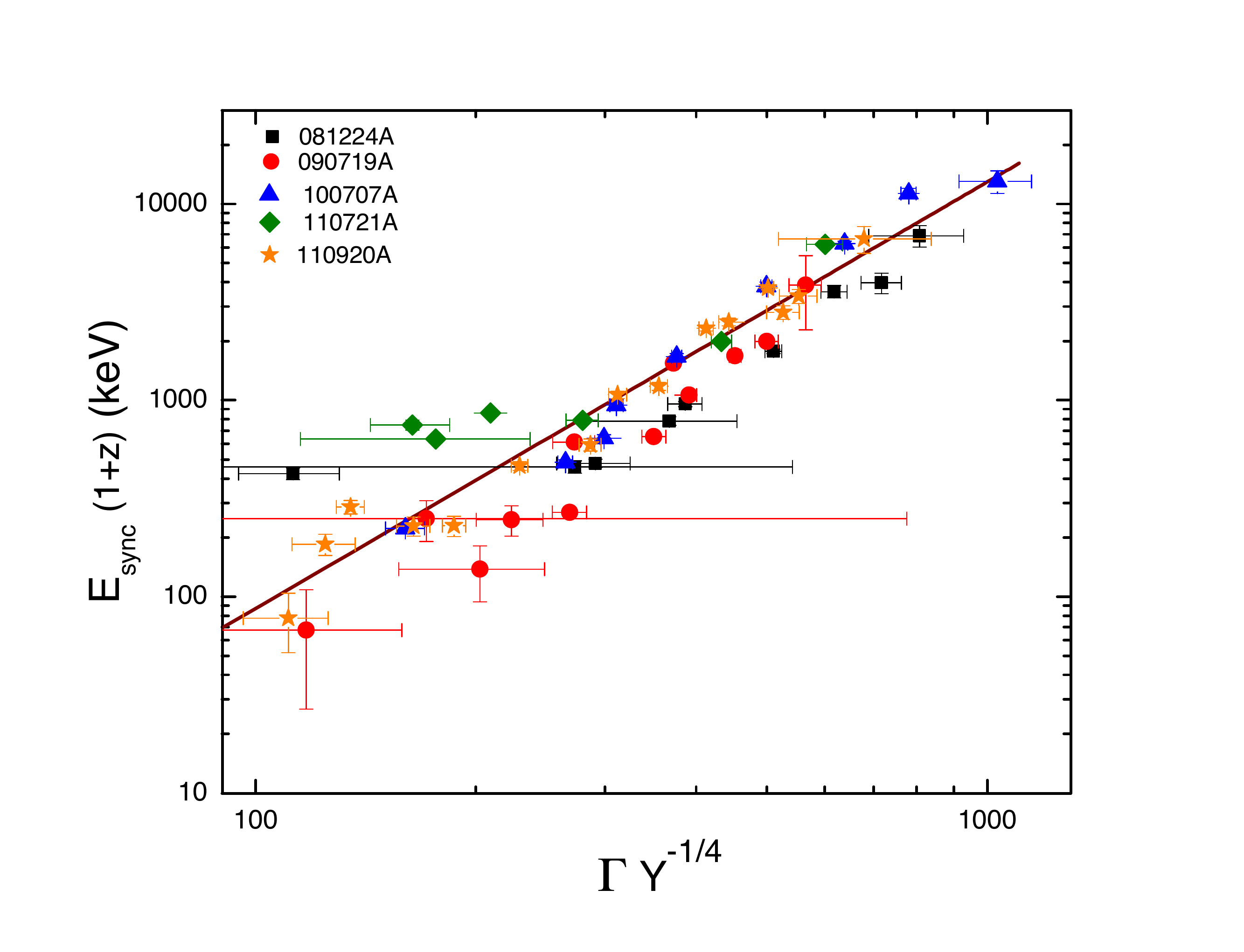}}
\caption{Peak energy of the synchrotron emission versus Lorentz factor correlation in the local frame (corrected for a redshift, $z = 2$). The correlation is given by $E_{\rm sync} (1+z) \propto \Gamma^{2.17 \pm 0.12}$. }
\label{gamma_Epk}
\end{center}
\end{figure}

\subsection{Constraints from the observations}
\label{sec:sync_constraints}

From the spectral fits we have determined the peak of the synchrotron component, $E_{\rm sync}$, {  and for the five burst in which the thermal component is detected, we have determined} the properties of the flow, for instance,  the Lorentz factor, $\Gamma$ and the photospheric radius, $r_{\rm ph}$  (see \S \ref{outflow_cal}). 

First, we 
%The synchrotron energy peak, given by the fit, $E_{\rm sync}$ is related to the Lorentz factor of the outflow as $E_{\rm sync} \propto \Gamma \gamma_{\rm el}^2 B$. 
 find that the synchrotron energy peak $[$corrected for a redshift, $z = 2$,  i.e $E_{\rm sync}\: (1+z)]$ shows a correlation with $\Gamma$ such that $E_{\rm sync} (1+z) \propto \Gamma^{2.17 \pm 0.12}$ (see Figure \ref{gamma_Epk}). 
%The correlations  between Luminosity and $E_{peak}$ as well as Luminosity and $\Gamma$. 
However, according to equation (\ref{E_c}), the peak energy $E_{\rm sync} (1+z) \propto \Gamma \gamma_{\rm el}^2 B$. Thus, the observed correlation suggests that $B \gamma_{\rm el}^2 $ is approximately proportional to the evolution in $\Gamma$. %which has a distinct evolution with time. 
This in turn suggests that both
$B$ as well as $\gamma_{\rm el}$ also evolve during the burst \citep{Beniamini2013,Uhm&Zhang2014}.  In the analysis we take, in each time-bin, an average value of the magnetic field and $\gamma_{\rm el}$ ({  see further discussion in \S \ref{sec:alt}}).

\begin{figure*}
\begin{center}
\resizebox{84mm}{!}{\includegraphics{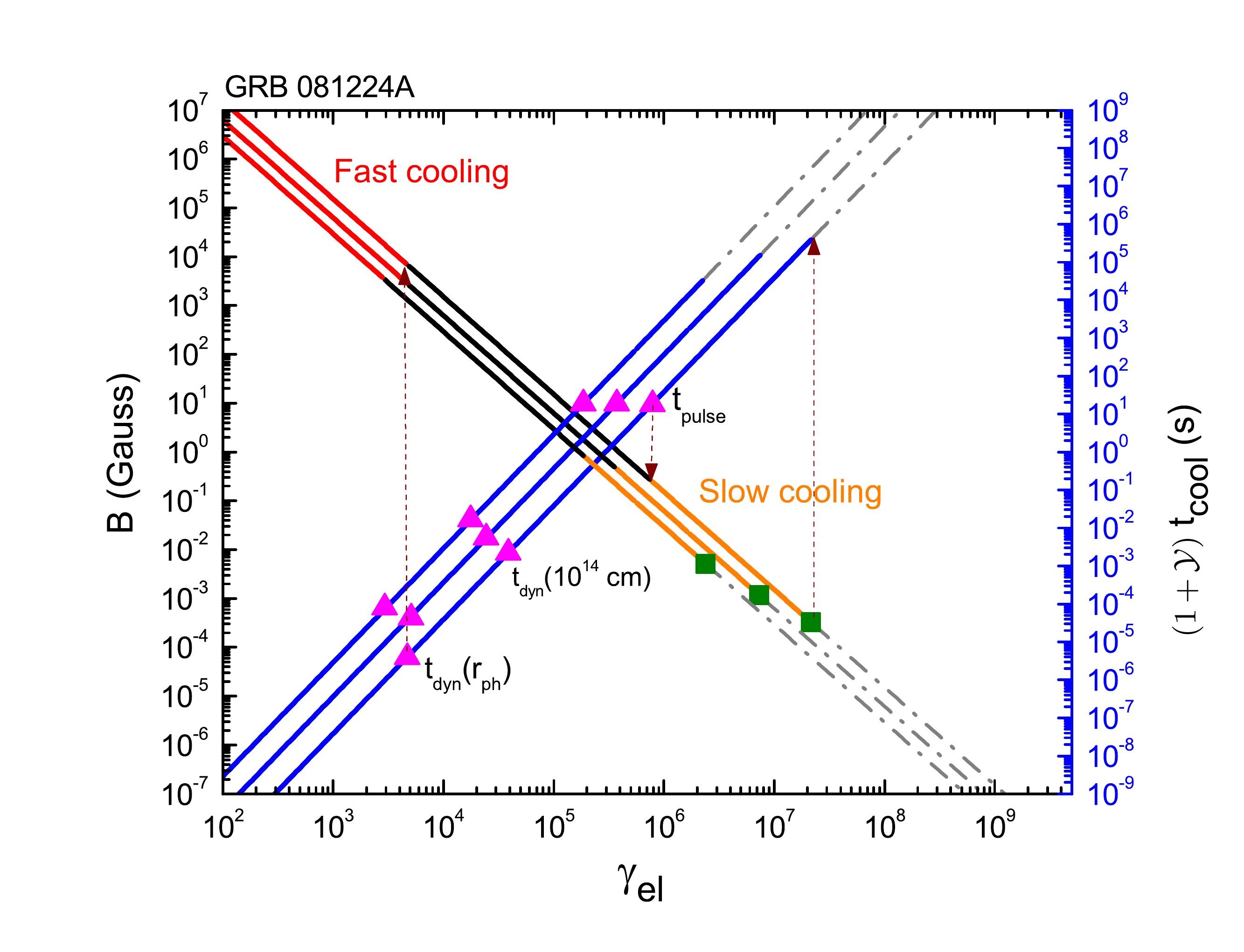}}
\resizebox{84mm}{!}{\includegraphics{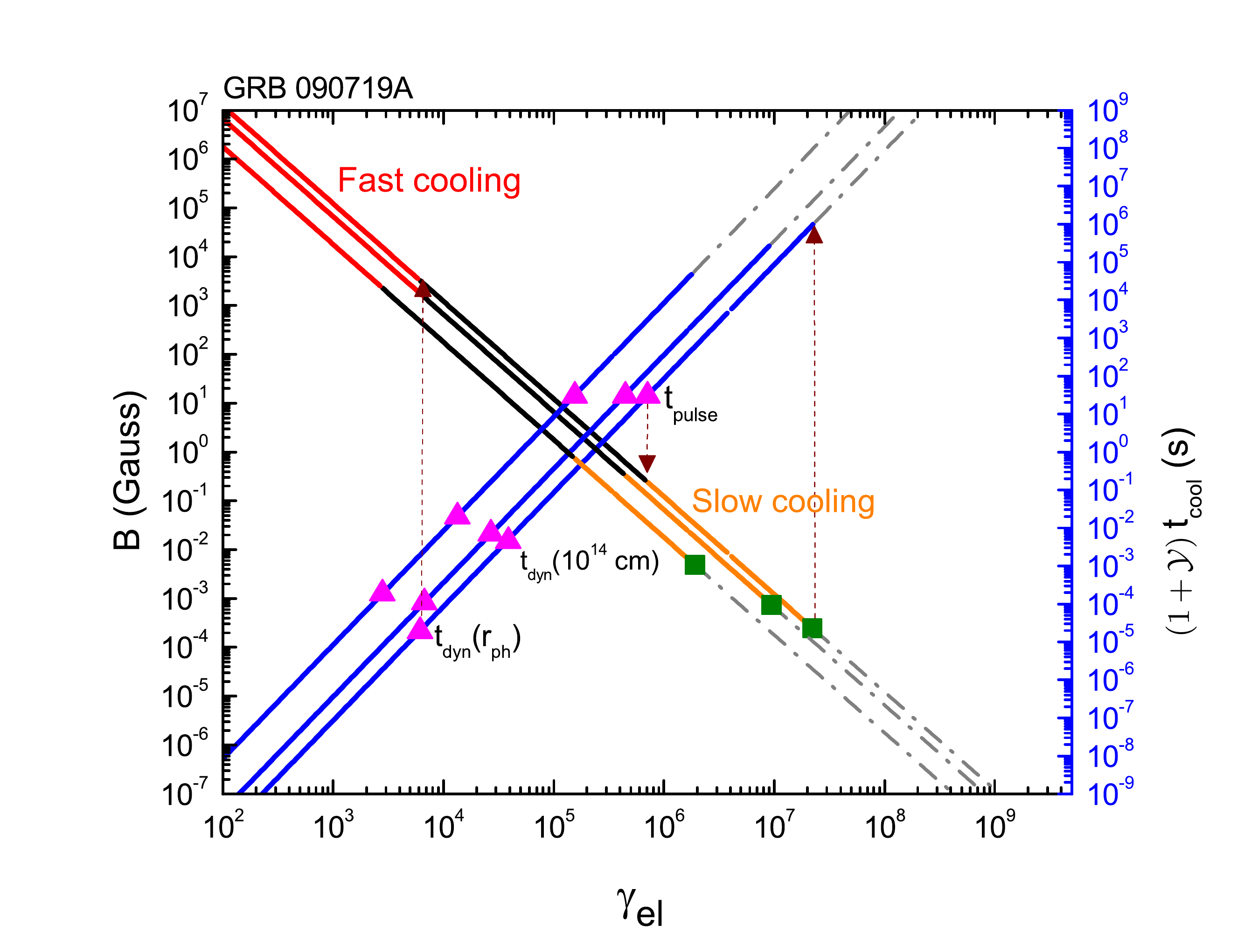}}
\resizebox{84mm}{!}{\includegraphics{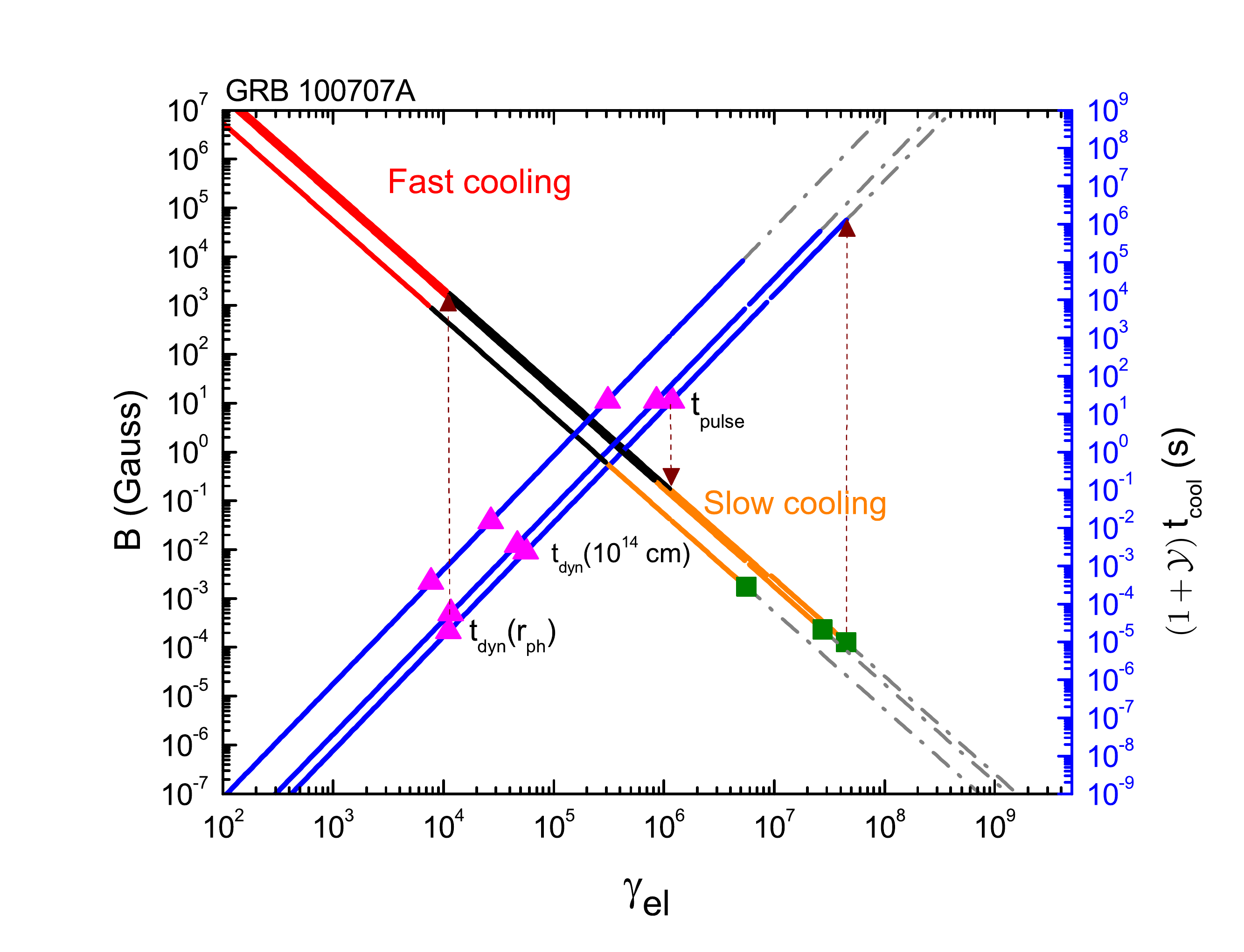}}
\resizebox{84mm}{!}{\includegraphics{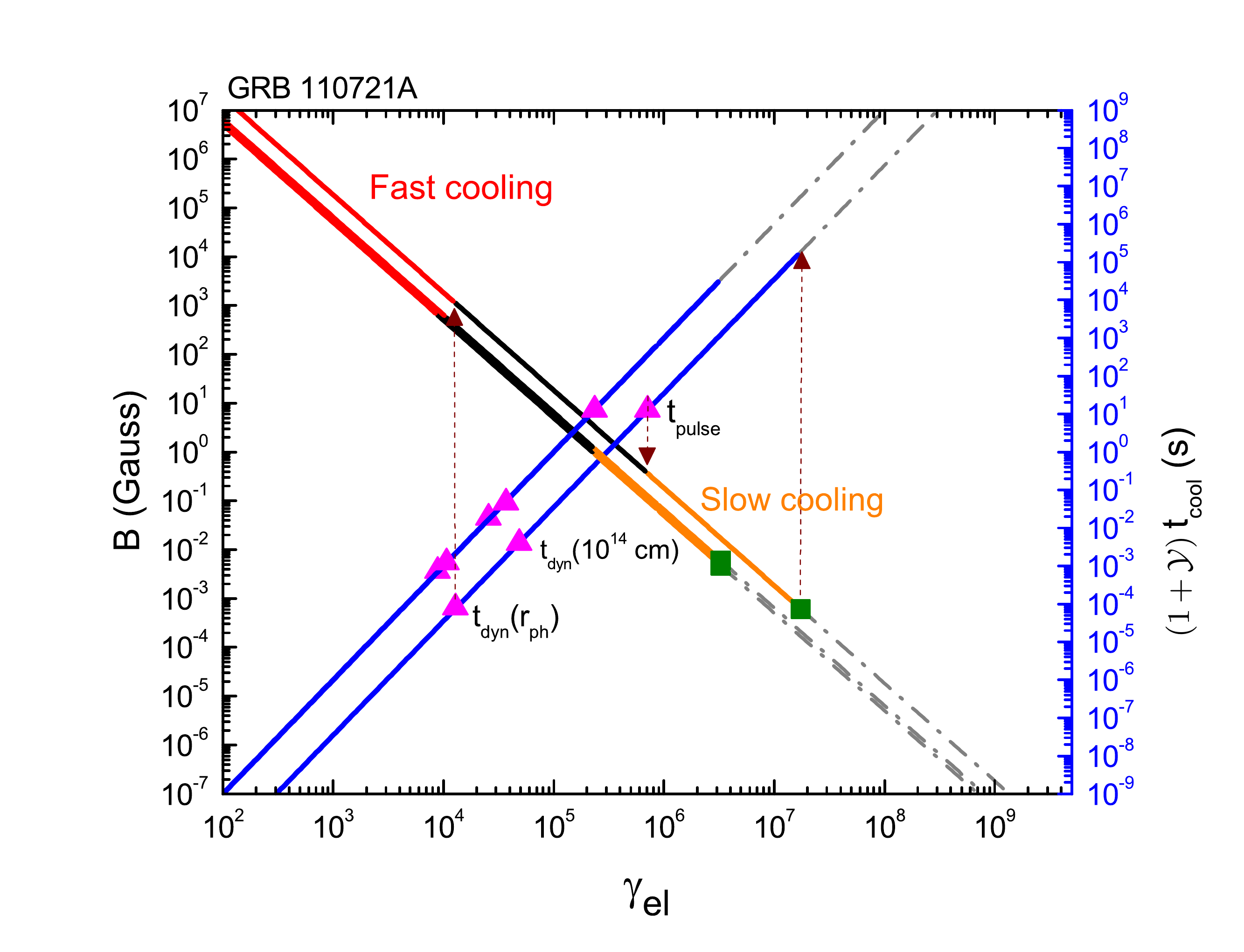}}
\resizebox{84mm}{!}{\includegraphics{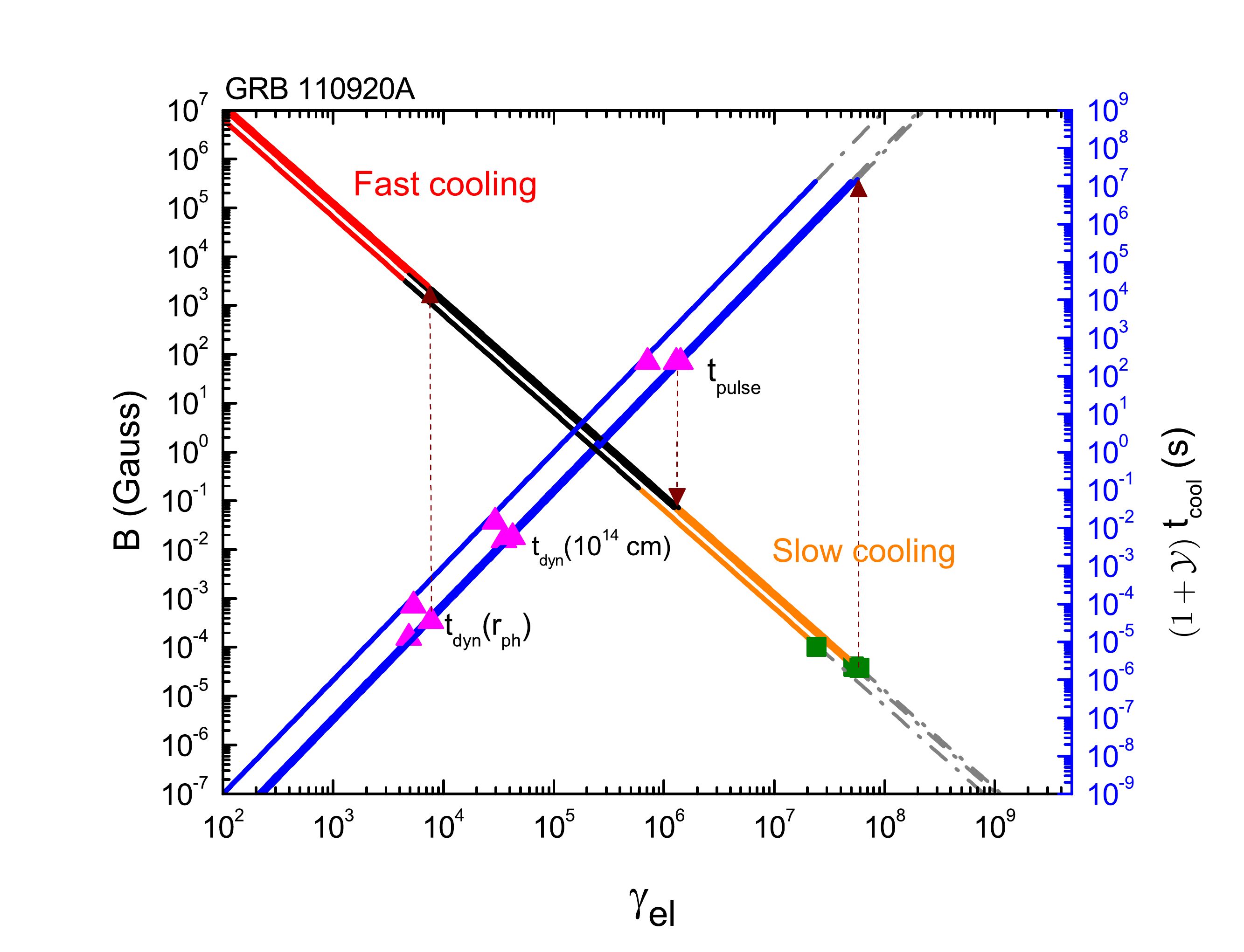}}
\caption{ Allowed relations between  $B$ and $\gamma_{\rm el}$. The black lines show the constraints obtained for $B\gamma_{\rm el}^2$ from the equation (\ref{B}) for three time bins: one before, one at and one after the peak of the light curve.  The blue lines show the dependence of $t_{\rm cool} $ on $\gamma_{\rm el}$. The dynamical time for different characteristic radii ($r_{\rm ph}$, $10^{14} \rm{cm}$) and $t_{\rm pulse}$ are shown with pink triangles. The red section of the lines shows the values of $B$ and $\gamma_{\rm el}$ that result in $t_{\rm cool} < t_{\rm dyn} (r_{\rm d})$ for all allowed values of $r_d > r_{\rm ph}$ i.e. the electrons always undergo fast cooling. The orange section of the lines shows the values of $B$ and $\gamma_{\rm el}$ that will always result in $t_{\rm cool} > t_{\rm pulse}$, i.e, electrons always undergo slow cooling. The black part of the lines represents the condition where the cooling of the electrons can be either fast or slow depending at which radius the dissipation occurs. $\tau_e \leq \tau_{\rm tot}$ gives a lower limit on $B$ and thereby a corresponding upper limit on the $\gamma_{\rm el}$, which is marked in green squares. The dash dot lines shown in grey on both the curves represent the forbidden parameter space of $B$, $\gamma_{\rm el}$ and $t_{\rm cool}$.  }
\label{fig:gamma_el_B}
\end{center}
\end{figure*}

Second, assuming that the properties of the  flow are the same at the photosphere and at the dissipation site, equation (\ref{E_c}) then gives  a constraint for the product $B \gamma_{\rm el}^2$ for every time bin in our observations: 
\begin{equation}
B \gamma_{\rm el}^2 = \frac{E_{\rm sync} (1+z) 4 \pi m_e c}{\Gamma 3 h q}. 
\label{B}
\end{equation}
\noindent
These are shown by the black lines in Figure \ref{fig:gamma_el_B}, where constraints obtained for three time bins are plotted: one before, one at and one after the peak photon flux in order to capture the time evolution. % {  Three times bins are chosen for each burst, which?}

Furthermore, from equations (\ref{tcool}) and (\ref{B}), we can choose to express  the cooling time as a function of $\gamma_{\rm el}$, which is plotted as blue lines in Figure \ref{fig:gamma_el_B} with the right hand (or blue) y-axis showing the cooling timescale. {  From equation (\ref{B}), we find low values of $B$, for large values of $\gamma_{\rm el}$. Since cooling timescale, $t_{\rm cool}$ is very sensitive to the changes in $B$ when these values are substituted in equation (\ref{tcool}), result in longer cooling time even for large values of $\gamma_{\rm el}$.}  %we can understand for what values of $B$ and $\gamma_{\rm el}$, electrons cool fast and slow with respect to the dynamical times. 
%Knowing $B$ and $\gamma_{\rm el}$ allows us to estimate the cooling time and thereby compare to the dynamical time for the various dissipation radius and conclude if the electrons would cool fast or slow. 
On the blue lines we mark  with pink triangles $t_{\rm pulse}$  and the dynamical time (eq. \ref{tdyn}) for $r_{\rm ph}$. In addition we plot the  dynamical time for $10^{14} \rm{cm}$, representing a time between these extremes. %The requirement  $t_{\rm cool} > t_{\rm dyn} ¨$ (slow cooling) corresponds to  $\gamma_{\rm el}$ with {  larger values.}

The red section of the lines show the values of $B$ and $\gamma_{\rm el}$ that result in $t_{\rm cool} < t_{\rm dyn} (r_{\rm ph})$  
for all allowed values of $r_d > r_{\rm ph}$. In other words, the electrons are always in the fast cooling regime.  
An upper limit of the dynamical time is given by the width of the pulse of the burst, $t_{\rm pulse}$ which corresponds to an upper limit of the allowed dissipation radius, $r_{d, \rm max} = 2 \Gamma^2 c t_{\rm pulse}$. Thus, for values of $B$ and $\gamma_{\rm el}$ depicted by the orange lines will always result in $t_{\rm cool} > t_{\rm pulse}$ (slow cooling) for the allowed values of $r_d$. %the bursts parameter values. %will always result in electrons being in the slow cooling regime. %cooling slowly with respect to the dynamical time for the allowed values of $r_d$. 
The values of $B$ and $\gamma_{\rm el}$ resulting in $t_{\rm dyn}(r_{\rm ph}) < t_{\rm cool} < t_{\rm pulse}$ (shown in black) can result in synchrotron emission from electrons either cooling fast or slow depending on where the dissipation occurs and what is the corresponding dynamical time. 

The maximal spectral flux at the peak frequency is given by 
\begin{equation}
\frac{F_{\rm sync}}{E_{\rm sync}} = \frac{\sigma_T \Gamma m_e c^2 B (1+z) \: N_e}{36 \pi q d_L^2}
\label{pk_flux}
\end{equation}
\noindent
 where $N_e = (8\pi \Gamma^2 \: r_d \: c \:t_{\rm dyn} \: \tau_e) /\sigma_T$ \citep{Beniamini2013}. 
Thus, substituting for $N_e$ in equation (\ref{pk_flux}), the opacity of the electrons radiating synchrotron emission is given by 
\begin{equation}
\tau_e = \frac{F_{\rm sync}}{E_{\rm sync}} \frac{9 q d_L^2 \: \tau_{\rm tot}}{2 \Gamma^3 m_e c^3 B\: r_{\rm ph} t_{\rm dyn} \: (1+z)}
\label{tau_e}
\end{equation}
\noindent
where $\tau_{\rm tot} = r_{\rm ph}/r_d$ is the opacity due to the electrons associated to the number of baryons in the outflow at $r_d$. $\tau_e$ cannot be larger than $\tau_{\rm tot}$, unless electron - positron pairs are created via dissipation, and $t_{\rm dyn}$ at the maximum can only be the observed width of the pulse, $t_{\rm pulse}$. This gives a lower limit on $B$ being between $10^{-3}$ and $10^{-4}$ G  and thereby an upper limit on $\gamma_{\rm el}$ lying between $\sim \rm{few} \, \times10^6$ and $10^8$ (marked in green squares in Figure \ref{fig:gamma_el_B}).  

%The observed synchrotron component is considered to be optically thin, as a result the radius of dissipation, $r_d$, lies between, $r_{\rm ph} < r_d < r_{\rm dec}$ where $r_{\rm dec}$ marks the radius at which the outflow crashes into the circumstellar ambient medium. 

%The fits to the data clearly rejects synchrotron emission from a cooled electron distribution, typically referred to as "fast cooled" synchrotron emission, instead  %Such spectra are too broad to be able to be fitted to the data even when a blackbody is added to the model.
%The observed spectrum 
Since the best spectral fit is for {  synchrotron emission from uncooled electron distribution}, the  general requirements %for slow cooling synchrotron emission 
are $B < 1000$ G and $\gamma_{\rm el} > 10^4$ for all bursts in the sample. In particular, to be fully in the slow cooling regime, $\gamma_{\rm el}$ should be larger than between $10^5$ and $10^6$. 

%The slow cooling synchrotron emission from impulsive heating raises the issue of the efficiency of the radiation which however is not an issue in case of fast cooling scenario.

%We find that for the synchrotron emission coming from fast cooling of electrons, the plausible parameter space of $B > 1G$ and $\gamma_{\rm el} < 10^6$. 
%However, 
%We find that only a part of this curve falls within the plausible parameter space of $\gamma_{\rm el}$ and $B$ and these are the permitted values of the $\gamma_{\rm el}$ and $B$ for the burst depending on $\Gamma$ and $R$. Interestingly, we note that for all the bursts in the sample the permitted values of $\gamma_{\rm el}$ and $B$ are nearly the same such that $B$ ranges between $10^{-0.8} - 10^{4.2}\: \rm G$ and $\gamma_{\rm el}$ ranges between $10^3 - 10^6$, see Figure \ref{fig:gamma_el_B}, which are nearly consistent with the predictions made in \cite{Beniamini2013}.  

\section{Decreasing $\Gamma$ and the internal shock model}

A common behaviour of the time-dependent analysis carried out here is that the Lorentz
factor decrease in time (\S 2.1).  This is a consequence of
the characteristic temporal evolution of the temperature and thermal
flux \citep{Ryde2004,Ryde2005,Pe'er2007, Ryde&Pe'er2009}.  Similar results were found in among others \cite{Ryde2010, Iyyani2013, Ghirlanda2013, Preece2014}. Such a behaviour can be ascribed to varying central engine properties (see also \cite{Iyyani2013}).
If this interpretation is correct, this behaviour excludes the possibility that the synchrotron
emission component is from internal shocks since for these to occur
later emitted shells of the jet have to catch up the preceding shells
in order to form shocks.  {  Therefore, in our interpretation the origin of synchrotron emission is from a forward shock. This is consistent with that all the bursts in sample are single pulses (Burgess et al. 2016, in prep.).}

%A common behaviour of the jets observed in this sample is that the Lorentz factor decreases (\S \ref{Gamma}). Similar results were found in \cite{Iyyani2013,Preece2014}. This behaviours excludes the possibility that the synchrotron emission component is from internal shocks since for these to occur later emitted shells of the jet have to catch up the preceding shells in order to form shocks. 

In the next sections, we discuss alternative assumptions that can be made in the derivation of the jet properties. In the estimations above we have assumed that the radiative efficiency and the magnetisation are constant over the burst. {  Moreover, we have assumed the flow to be baryon-dominated and 
have neglected high-latitude effects.} Below, we investigate {  and discuss} what happens if we relax these assumption.

%\subsection{Alternative Scenario for outflow}
\subsection{Radiative efficiency: $Y$ parameter }
 \label{Y}
The radiative efficiency of the burst is given by $Y^{-1}$ (eq. \ref{Y_rad}). %, the ratio of total burst energy to the observed $\gamma$ -ray energy and 
In the estimations of $\Gamma$ above, there is  a dependence on the efficiency, $\Gamma \propto Y^{1/4}$ (eq. \ref{Gamma_eq}). 
Therefore,  the determined evolution in $\Gamma(t)$ could, fully or in part, be attributed to corresponding variations in $Y(t)$. Estimations of $Y$ can be made from afterglow measurements \citep{Racusin2011, Wygoda2015}. Unfortunately, for none of the bursts in the sample, afterglows have been observed.

We therefore study the limiting case where we assume  that $\Gamma$ remains a constant throughout the burst and determine what requirements then are set on $Y$. Since $Y$ has to be greater than unity, the assumed constant value of $\Gamma$ has to be larger or equal to the highest value of the currently estimated $\Gamma$ . %($Y$ is initially 1)}  
We now choose a value of constant $\Gamma$, equal to the highest value currently estimated. We then find the corresponding estimated value of  $Y$ to increase with time from nearly $1$ to $1000$.  

\begin{figure}
\begin{center}
%\resizebox{84mm}{!}{\includegraphics{081224A_Y_r0.eps}}
%\resizebox{84mm}{!}{\includegraphics{090719A_Y_r0.eps}}
\resizebox{84mm}{!}{\includegraphics{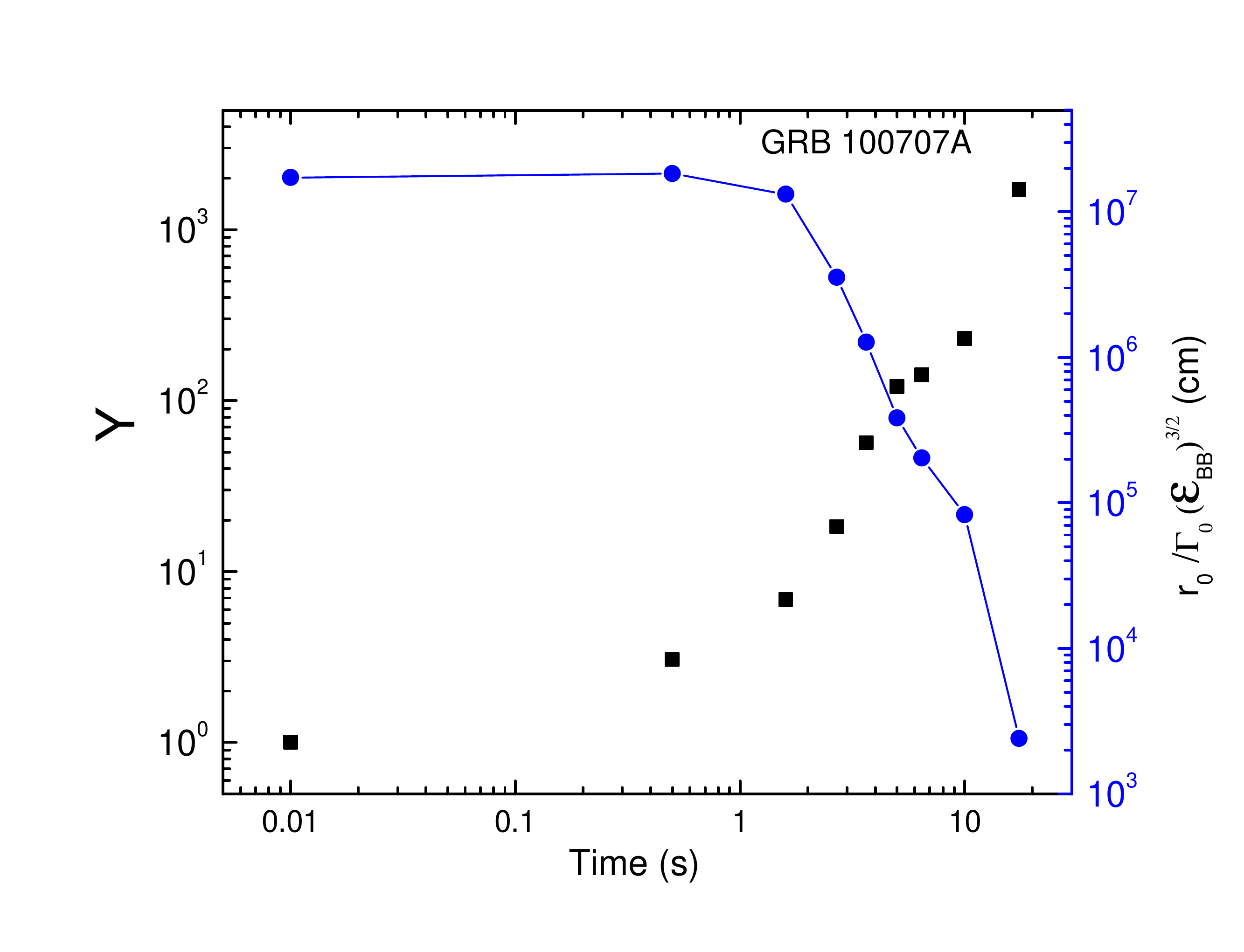}}
%\resizebox{84mm}{!}{\includegraphics{110721A_Y_r0.eps}}
%\resizebox{84mm}{!}{\includegraphics{110920A_Y_r0.eps}}
\caption{In order to keep the Lorentz factor a constant throughout the burst requires $Y$ (black/ squares) to increase with time. %Along with that, we estimate 
This in turn implies that $r_0$ (blue/circles with line) has to decrease and reach unreasonably low values.}
\label{Y_r0}
\end{center}
\end{figure}

This evolution of $Y(t)$, will affect  the value of $r_0$ as well. We find that in all bursts then $r_0$ decreases from nearly $10^{8}$ to $10^3$ cm. As an example, the case of GRB100707A is shown in Figure \ref{Y_r0}.
These values should be compared to the Schwarzschild radius of the central black hole, which is of  the order of $10^{6.5}$ cm (for black hole of masses $5 -10\: M_{\sun}$ \cite{Paczynski1998}). Since the inferred values are smaller than the Schwarzschild radii, such an  evolution of $Y(t)$ has to be rejected. Variations in the radiative efficiency can thus not account for the observed decrease in $\Gamma$. In addition to this, recently \cite{Wygoda2015} found that there is relative small scatter in the estimates of the radiative efficiency of bursts and find an average value of $Y \sim 2$. This suggests that large variations in $Y$ within individual bursts are unlikely as well. 

\subsection{Magnetisation parameter, $\sigma_0$}
{  The analysis above assumes that the magnetic field is subdominant. However, we have no way of directly measuring the magnetic field. It was shown by \cite{Zhang&Pe'er2009} that if the magnetic field is dominant then the photospheric component is suppressed, which thus could explain the 4 bursts with no or only a weak thermal component. On the other hand, as argued above, the presence of a strong thermal component with a flux $\sim 40 \%$ of the observed total flux (see Figure \ref{fig:ratio}) and at an energy well below 1 MeV, suggest that the outflow in these bursts are baryonic dominated.  Still, for these bursts the outflow may be moderately magnetised. To investigate the effect of such a magnetisation on the derived parameters, we consider the hybrid model outlined in \cite{Gao&Zhang2015} (see also \citet{Iyyani2013}), with a dominant baryonic component such that the outflow is predominantly accelerated by the thermal pressure. } 
{ Moreover, for such a baryonically dominated outflow, there is a large paramerter space region in which the photosphere radius is above the coasting radius. In the following, we assume that this is the case, which corresponds to region III discussed in Gao \& Zhang (2015).}

%Recently, \cite{Gao&Zhang2014} have considered 
%In a hybrid model for the GRBs  \citep{Gao&Zhang2015, Iyyani2013} 

In such a case, the total burst luminosity, 
\begin{equation}
L_0 = L_h + L_c
\label{L_0}
\end{equation}
\noindent
where $L_h$ is the hot component (fireball, characterised by $\eta \equiv L_h/\dot M c^2$) and $L_c$ is the cold component (Poynting flux, characterised by $\sigma_0$) of the burst. 
The magnetisation parameter, $\sigma_0$ is defined as 
\begin{equation}
\sigma_0 = \frac{L_c}{L_h}
\label{GZ_1}
\end{equation}
\noindent
The conditions  $\sigma_0 \ll 1$ and $\eta \gg 1$ result in a pure fireball while $\sigma_0 \gg 1$ results in a highly magnetised outflow.  
{  In this scenario, considering that there is no magnetic dissipation taking place below the photosphere, the outflow parameters are obtained as follows: } 

%\subsubsection{Photosphere in the coasting phase}
%The parameter ${\cal{R}}$ is related to the transverse size of the observed photosphere by 
%\begin{equation}
%{\cal{R}} =1.06  \frac{(1+z)^2}{d_L} \frac{r_{\rm ph}}{\Gamma_{\rm ph}}
%\label{GZ_3}
%\end{equation}
%\noindent
The Lorentz factor at the photosphere, $\Gamma_{\rm ph}$ is given by  
\begin{equation}
\Gamma_{\rm ph} %= \Gamma_c  
=  \eta(1+\sigma_0),
\label{Gamma_c}
\end{equation}
\noindent
see equation (12) in \cite{Gao&Zhang2015}. %{  assuming that the magnetisation at the photosphere is low, which is consistent with the observation of strong blackbody in the spectrum \citep{Zhang&Peer2009}.}
The photospheric radius, $r_{\rm ph}$, is given by 
\begin{equation}
r_{\rm ph} = \frac{L_0 \sigma_T}{8 \pi m_p c^3 \Gamma_{\rm ph}^2 \eta(1+\sigma_0) },
\label{GZ_4}
\end{equation}
\noindent
see equation (18) in  \cite{Gao&Zhang2015}.
Substituting equation (\ref{GZ_4})  in equation (\ref{R}) for ${\cal{R}}$ (see \cite{Pe'er2007})
%\begin{equation}
%{\cal{R}} = \phi  \frac{(1+z)^2}{d_L} \frac{r_{\rm ph}}{\Gamma_{\rm ph}}
%\label{R}
%\end{equation}
%\noindent
%where $\phi$ is a factor of the order of unity, 
 gives the estimate of $\eta$ of the burst,  
\begin{equation}
\eta = \left[\frac{L_0 \sigma_T \phi (1+z)^2}{8\pi m_p c^3 d_L (1+\sigma_0)^4 {\cal{R}}}\right]^{1/4}
\label{GZ_5}
\end{equation}
for a given value of $\sigma_0$ {  and $\phi$ is a factor of order of unity}. 
Once knowing $\eta$ and substituting in equation (\ref{Gamma_c}) gives the estimate of $\Gamma_{\rm ph}$ and thereby $r_{\rm ph}$. 
%\begin{equation}
%\Gamma_{\rm ph} = \eta (1+\sigma_0)
%\label{GZ_6}
%\end{equation}
%\noindent
{  It is worth noting that in this assumed scenario, the estimates of $\Gamma_{\rm ph}$ and $r_{\rm ph}$ have no dependence on the magnetisation parameter, $\sigma_0$ after the substitution for $\eta$, and the expressions obtained are equivalent to the ones obtained in \cite{Pe'er2007}. }
%Thus, the estimate of the photospheric radius in this regime of coasting phase also has no dependence on $\sigma_0$ and would be equivalent to the one obtained in \cite{Pe'er2007}. 
The difference to be noted is that in this scenario $\Gamma_{\rm ph} \neq \eta$ which was otherwise the case in \cite{Pe'er2007}.   

%The saturation radius, $r_s$ is given by 
%\begin{equation}
%r_s = r_{ra} \left(\frac{\Gamma_c}{\Gamma_{ra}}\right)^{1/\delta}
%\label{GZ_7}
%\end{equation} 
%where we assume $\delta = 1/3$. 
%To estimate $r_{s}$ we need to estimate $r_{ra}$.%which can be estimated from the observed temperature of the blackbody relation in this regime. 
% However, we cannot estimate $r_{ra}$. 

The observed blackbody temperature, $T$, at the photosphere (see equation 22 in \cite{Gao&Zhang2015}) is given by
\begin{equation}
T = \frac{\zeta\: \Gamma_{\rm ph}}{(1+z)} T_0 \left(\frac{r_{\rm ra}}{r_0}\right)^{-1} \left(\frac{r_s}{r_{\rm ra}}\right)^{-(2+\delta)/3} \left(\frac{r_{\rm ph}}{r_s}\right)^{-2/3}
\label{T}
\end{equation}
\noindent
where $\zeta$ is a factor of order of unity. The acceleration is initially mediated by the photons, at $r < r_{\rm ra}$, and at larger radii the acceleration is dominated by the reconnection by the magnetic field, resulting in $\Gamma \propto r^{\delta}$ ($\delta = 1/3$)  until the saturation radius $r_s$.  
\begin{equation}
T_0 = \left(\frac{L_0}{4\pi r_0^2 ca (1+\sigma_0)}\right)^{1/4}
\end{equation} 
\noindent
where $a$ is the radiation constant and 
\begin{equation}
r_s = r_{\rm ra} \left(\frac{\Gamma_c}{\Gamma_{\rm ra}}\right)^{1/\delta}
\end{equation}
\noindent
where $\Gamma_{\rm ra} = r_{\rm ra}/r_0$. %, see \cite{Gao&Zhang2015}. 
% where, $T_0$ is the plasma temperature at the nozzle radius, $r_0$, we are able to estimate the nozzle radius of the jet, $r_0$, which is given by 
%\begin{equation}
%T_{\rm ob} = \frac{1.48 \Gamma_{\rm ph}}{(1+z)} T_0 \left(\frac{r_{\rm ra}}{r_0}\right)^{-1} \left(\frac{r_s}{r_{ra}}\right)^{-(2+\delta)/3} \left(\frac{r_{ph}}{r_s}\right)^{-2/3}
%\end{equation}
% \noindent 
% where $T_0 = \left(\frac{L_0}{4\pi r_0^2 ca (1+\sigma_0)}\right)^{1/4}$;  $\delta = 1/3$ and $\Gamma_{ra} = r_{ra}/r_0$. This gives the expression 
% \begin{equation}
% T_{ob} = \frac{1.48}{(1+z)} T_0 \left(\frac{r_{ph}}{\Gamma_{\rm ph}}\right)^{-2/3} r_0^{2/3}
% \end{equation}
% \noindent
% which is similar to the expression in \cite{Pe'er2007}. 
{  One can use the expression in equation (\ref{T}), and after some algebra,  solve for the nozzle radius of the jet, $r_0$:}
 \begin{equation}
 r_0 =   \psi \: \frac{{\cal{R}} \: d_L}{(1+z)^2} \left(\frac{F_{\rm BB}}{F}\right)^{3/2} \left(\frac{1+\sigma_0}{ Y }\right)^{3/2}
 \end{equation}
 \noindent
 where $\psi$ is a factor of order of unity. 
% which is the same equation discussed in \cite{Iyyani2013} when we consider a magnetised central engine with no subphotospheric dissipation. 
 
 In summary, if we assume a value for $\sigma_0$, we can estimate the outflow parameters $r_0, \eta $ and thereby $\Gamma_{\rm ph}$. However, in this scenario we cannot estimate $r_s$ as 
% $r_s$ is given by 
%\begin{equation}
%r_s = r_{ra} \left(\frac{\Gamma_c}{\Gamma_{ra}}\right)^{1/\delta}
%\label{GZ_7}
%\end{equation} 
%where %$\delta = 1/3$ \citep{Drenkhahn&Spruit2002}, and 
$r_{\rm ra}$ is unknown. 

 \begin{figure*}
\begin{center}
\resizebox{84mm}{!}{\includegraphics{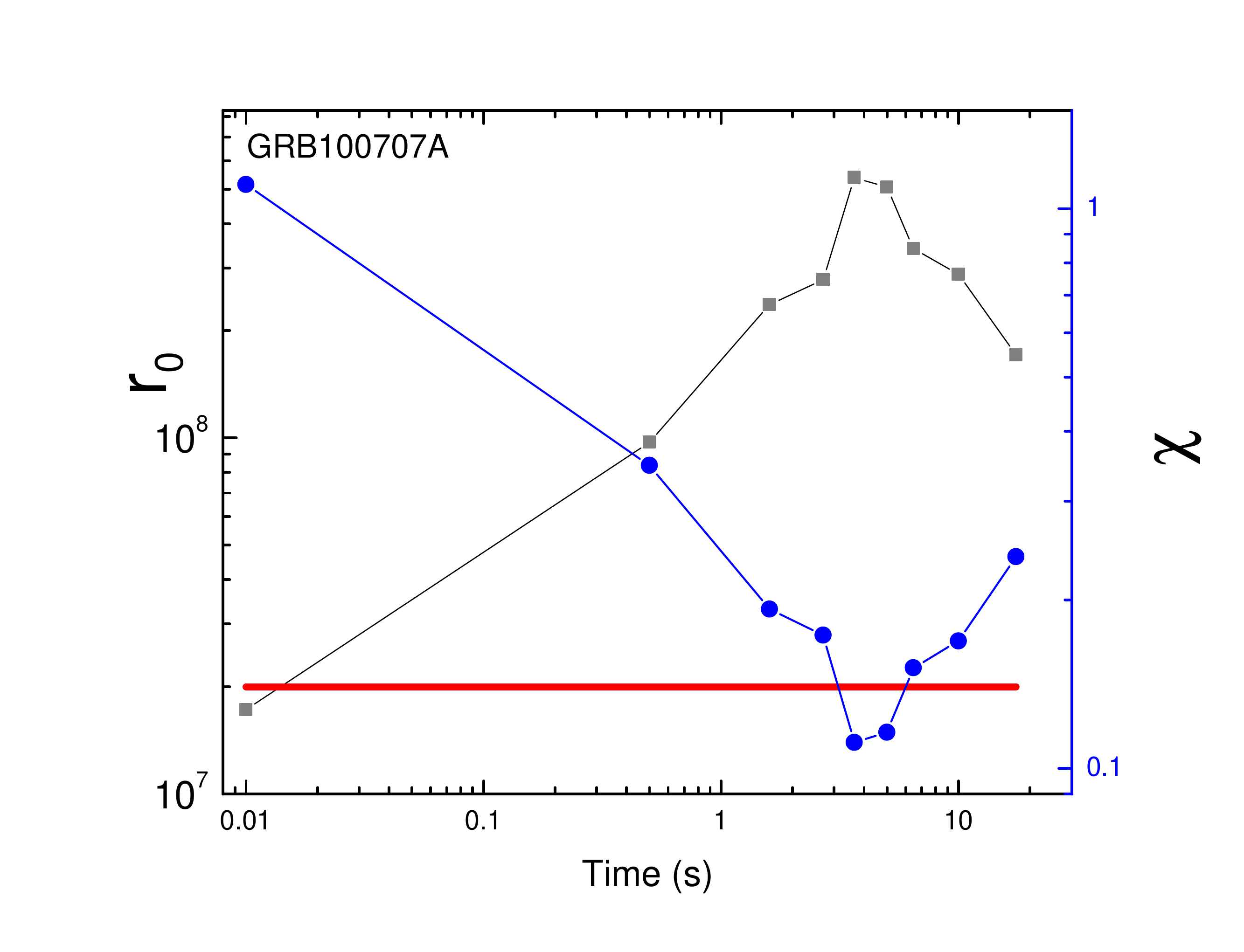}}
\resizebox{84mm}{!}{\includegraphics{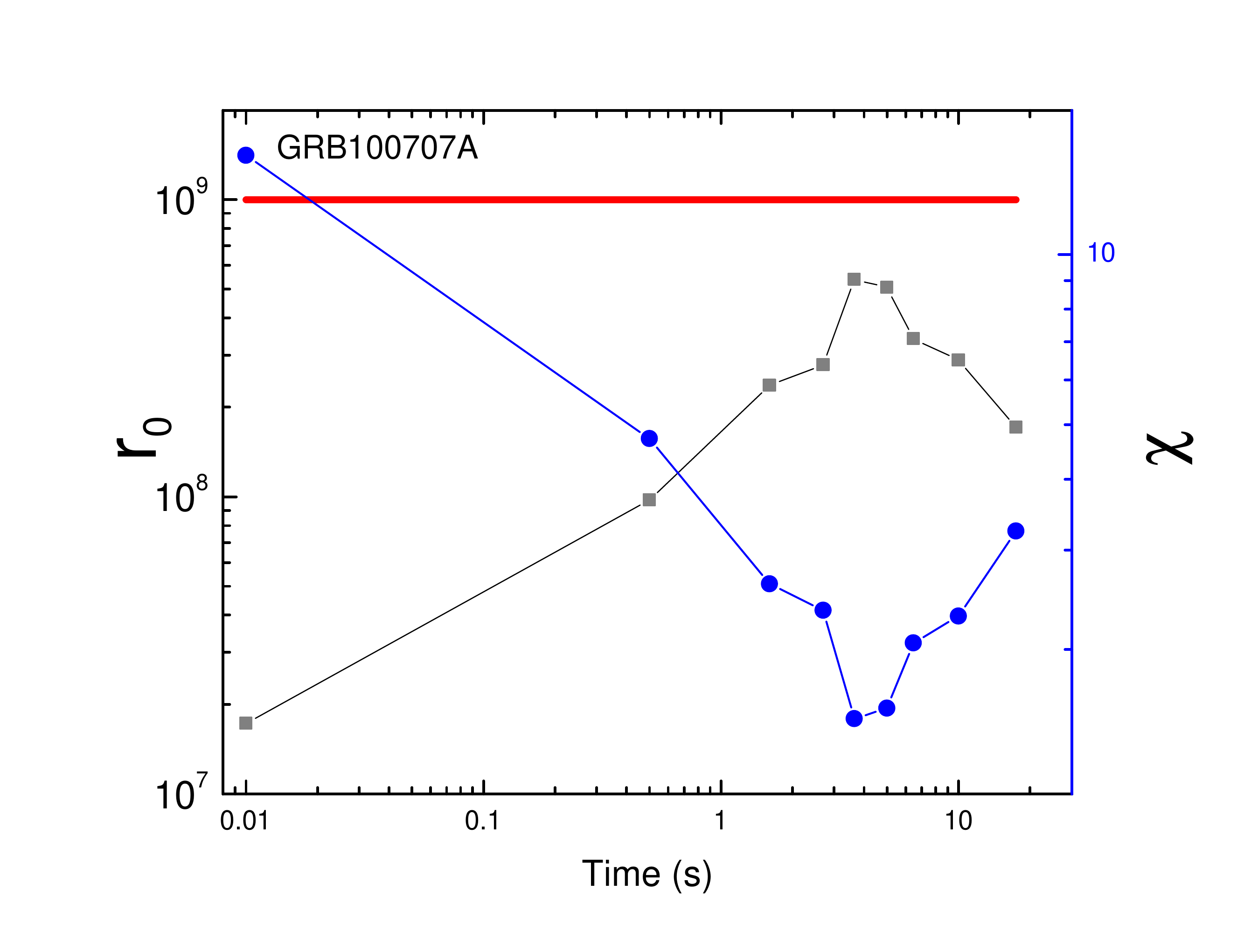}}
\caption{If $r_0$ is assumed to be a constant throughout the burst duration at  $r_0 \sim 2 \times 10^{7} \rm cm$, marked in red solid line (left hand panel) or $r_0 \sim 10^{9} \rm cm$, marked in red solid line (right hand panel), we find $\chi \equiv (1+\sigma_0)/\epsilon_{\rm BB} Y$, which gives a measure of the magnetisation of the outflow, to evolve as shown (blue circles with line). The otherwise deduced evolution of $r_0$ is shown in shade (grey squares with line). }
\label{r0_chi}
\end{center}
\end{figure*}

 A general assumption that is made in the literature is that $r_0$ is the size of the central engine and remains constant through out the burst duration.  If we assume $r_0$ to be a constant, we explore the variation that is possible in the unknown quantity $(1+\sigma_0)/\epsilon_{\rm BB} Y$ which we parameterise as $\chi$.  Figure \ref{r0_chi}a, for the case of GRB100707A, shows that when $r_0 = 2 \times 10^7$ cm throughout the burst, the $\chi$ varies between values $1$ and $0.1$. This clearly rules out any possibility of having $\sigma_0$ much larger than 1 (if $Y \sim 1$ ) which means the burst is weakly magnetised. Figure \ref{r0_chi}b shows that when we assume $r_0 = 10^9$ cm, we find $\chi$ to vary between $15$ and $1.5$. This implies that the $\sigma_0 > 1$; however since it is not $\sigma_0 \gg 1$, in such a case the burst is only moderately magnetised. %The estimates are valid provided $Y \sim 1$ as is expected in the scenario of ICMART model. 
This is consistent with the fact that we observe a strong blackbody component in the spectrum. 

%Thus, even when considering a scenario where the central engine is magnetised, if there is no subphotospheric dissipation, the outflow parameters obtained by the methodology in \cite{Pe'er2007} are very robust.   

% In such a hybrid model, $Y$ (equation \ref{Y_rad}) is given by %$Y = (L_b+L_p)/L_{\rm obs}$ which gives 
% \begin{equation}
% Y = \frac{L_b}{L_{\rm obs}}(1+\sigma_0).
% \end{equation}
% \noindent
%using equations (\ref{L_0}) and (\ref{GZ_1}). If we assume all of the fireball energy ($L_b$) is fully dissipated and thereby observed i.e $L_b \sim L_{\rm obs}$, then we find $Y \sim (1+\sigma_0)$. In such cases, the estimate of $Y$ gives a direct estimate of the magnetisation of the central engine. In the scenario discussed in \S \ref{Y}, where $\Gamma$ remains constant or increases with time, requires $Y$ to increase with time. %dramatically with time. This 
%From the above discussion, this requires that the magnetisation of the burst increases with time. 
%However, such a behaviour of $\sigma_0$ is not what is typically expected over a single pulse emission (where the peak energy decreases with time), in models invoking magnetic turbulence in the outflow, for instance \cite{Zhang&Yan2011}.%or magnetar models \citep{Metzger2010}.  
 %Otherwise, the estimate of $Y$ relates to the amount of energy that has not been dissipated and thereby not observed  (fraction of $L_0$), neglecting any loss of energy from the system. 

{ In conclusion, for the bursts with strong thermal components, the decreasing Lorentz factor can neither be due to a varying radiative efficiency nor a varying magnetisation of the jet {(assuming the photosphere radius is above the coasting radius)}. 
For the other bursts we have very poor constraints on the magnetic field. For  the three bursts where we did not have any clear detection of thermal component  we did not carry out any analysis, while for GRB110721A we perform a similar analysis to the others.  This can be justified by the fact that a subdominant thermal component does not necessary mean that the magnetisation is high. We point out that an alternative analysis for GRB110721A has been carried out by \cite{Gao&Zhang2015}, see (\S \ref{sec:Poynting}).}

\section{Discussion}

\subsection{Electron acceleration and magnetic field strength}
\label{sec:alt}

Within the interpretation of the spectrum presented here a large part of the spectrum is due to synchrotron emission. 
Since the electron distribution appears to be in the slow cooling regime, this sets strong limitations on the typical energy of the radiating electrons, for instance the minimum electron Lorentz factor, $\gamma_{\rm el} \equiv \gamma_{\rm min}$  should be larger than $10^5$ to $10^6$ (see also \cite{Beniamini2013}).  Such values are much larger than the typically assumed values, up to  $\gamma_{\rm min} \sim m_{\rm p}/m_{\rm e} = 1836$,  expected in various internal shock scenarios \citep{Bosnjak2009}, {  but are more typical for what is expected for forward shocks, see \S \ref{alter_scenario}}. 

%%%%%%%%%
%%%%%%%%%%
%Moreover, the low values inferred on the magnetic field (Figure \ref{fig:gamma_el_B}) means that a large fraction of the energy is stored in the electrons, which in turn would suggest that an inverse Compton component should be prominent at high energies \citep{Piran2009}.  {  XXXX However, it should be noted that at such high electron Lorentz factors, electrons at $\gamma_{\rm min}$ are expected to be in the Klein-Nishina regime, thereby suppressing any synchrotron self-Compton component (SSC).} %This however severely challenges the energy budget {\it DAMIENS comment here} \citep{Kumar&McMahon2008}.

One possibility is that only a small fraction of the electrons receive the dissipated energy and are accelerated forming a power law distribution with $\gamma_{\rm min} \gg m_{\rm p}/m_{\rm e}$ disconnected to the thermal distribution of electrons \citep{Daigne&Mochkovitch1998, Beniamini2013}. It is only the energetic electrons which radiate and the thermal electrons do not participate in the emission. 
%\subsection{Slow cooling or reheating?}
%\label{sec:reheating}
%The requirement to get a spectrum that matches the observations is therefore the the emitting electrons attain a very large, typical Lorentz factor, $\gamma_{\rm el}$.
%Thus, we get a plausible parameter space of $B \sim 10^{-4} - 10^3 \: G$ and $\gamma_{\rm el} \: \sim 10^4 - 10^7$ that can result in synchrotron emission from electrons cooling slowly. 
However, simulations of collisionless shocks show that  %{  Check spitskovski and other simulations} 
most of the electrons are accelerated and form a Maxwellian distribution with only a small contribution from a power law tail \citep{Spitkovsky2008}. %In such a case, the electron Lorenz factor of a typical electron should have the value of the order of $\gamma_{\rm min} \sim \rm{m_p}/m_{\rm. e} \sim 1836$ \citep{Bosnjak2009}, which strongly contradicts the observations. A possible way out of this is to assume 
 %and thus, the efficiency problem still remains. 
%Simulations of shocks needs yet to show under which conditions such a distribution can be formed. 
%The large values of $\gamma_{\rm el}$ suggests that only a fraction of the total electrons are accelerated \cite{Daigne&Mochkovitch1998, Bykov&Meszaros1996}. This suggests that the electron distribution is a Maxwellian with a disconnected power law. However, it is unknown if such an electron distribution can be achieved in shocks.  

%Along with that, the electrons in the slow cooling regime do not radiate all of their energy which leads to a decrease in the radiative efficiency of the synchrotron component

It can also be imagined that  the electrons are in the fast cooling regime, however their distribution is maintained through a balance between heating and cooling. Since the flow is baryonic (based on the observation of the BB component) the heating is mainly assumed to be due to shocks. Baryonic shocks, however, cannot maintain such a balance since the shocked particles will rapidly leave the shocked zone and cool undisturbed (\cite{Ghisellini&Celotti1999}, see also \cite{Kumar&McMahon2008}). 
On the other hand, in a scenario suggested by \cite{Pilla&Loeb1998,Medvedev&Loeb1999}, there can be an extended shock scenario where the shocked region is extended over a large volume due to Rayleigh -Taylor instability (see also \cite{Duffell&MacFadyen2014}). In such a case, the mean energy of electrons would be dictated by the balance between heating and cooling and this would result in values of $\gamma_{\rm min} < m_p/m_e$. This in turn from equation (\ref{B}) requires that the magnetic field, $B > 10^4 \rm G$ (see Figure \ref{fig:gamma_el_B}). %  The instability inevitably produces a minimum $\epsilon_{\rm B}$ i.e fraction of dissipated energy that goes into magnetic field, to be $\sim 10^{-5}$, which for a luminosity $L_0 = 10^{52} \rm erg$ and radius $r_d = 10^{12} \rm cm$, gives an upper limit for $B \sim 10^6 \rm G$.  }

Another way of relaxing the condition of slow cooling and still  maintaining the observed electron distribution is a marginally fast cooling scenario \citep{Daigne2011}. Here the characteristic Lorentz factor of electrons, when the cooling time is equal to the dynamical time, $\gamma_c \leq \gamma_{\rm min}$, which is opposed to the requirement of  $\gamma_c \ll \gamma_{\rm min}$ for the   fast cooling regime. The spectral peak is formed at $\nu_{\rm c, eff}$ 
%and $\Gamma_{c,eff} < \Gamma_c$ due to inverse Compton scattering. 
such that the photon index below $\nu_{\rm c, eff}$ is $-2/3$ (slow cooled) %. Thus, when $\nu_{c, eff} \sim \nu_m$, the observed photon index may be fit by $-2/3$ 
even when in the fast cooling regime. 
%This also helps to solve the issue regarding the low radiation efficiency in case of a slow cooled synchrotron emission. 
However, such a cooling tends to occur at large radii where the magnetic field, $B$ is low and the outflow has a large Lorentz factor. 
In our analysis, we find $B$ to be large for the case where fast cooling condition is satisfied (i.e $t_{\rm cool} < t_{\rm dyn}$) and thus, suggests that the cooling may not be marginally fast cooling. This also requires fine tuning in order to get $\gamma_c \leq \gamma_{\rm min}$.

{  In the spectral fits discussed in this paper the synchrotron component is associated with a certain value of the magnetic 
field strength.  The underlying assumption is that the magnetic field is constant in the emission region and does neither take into account inhomogeneities of the field strength nor the possibility of an evolving magnetic field or of an evolving injection rate of electrons.}
A possibility to address the large values of $\gamma_{\rm min}$ is therefore a scenario where the plasma is inhomogenous such that the magnetic field strength varies in different regions of the flow and the electrons emit radiation only when in the regions of strong magnetic field \citep{Pe'er&Zhang2006, Beniamini&Piran2014}. In such a scenario, the electrons can escape the emitting regions before they have the time to cool significantly and thus remain uncooled. Since only a small fraction of the energy stored in the electrons is radiated away, the energy budget is severely strained. 

{  \citet{Uhm&Zhang2014} suggested a model in which the emission region streams outwards in an expanding jet and therefore  the magnetic field strength in the emission region decreases with radius $[ B(r)\propto r^{-b} ]$, with $b \sim$ 1 -- 1.5. %For a constant injection rate (q=0) the spectra are broader than ob served.
The great advantage of such a model, compared to traditional synchrotron models,  is that it has a physical prescription  to explain the observed curvature and shape of the spectra in terms of varying magnetic field.
\cite{Zhang2015} uses such a model to fit the spectra of GRB130606B and find that it can fit the data well, provided a large value of $\gamma_{  min} \sim 10^5$. In addition, they find that, in order to explain the spectra with {  the Band function} $\alpha \sim - 0.8$,  a rapid increase in the electron injection rate is needed;  $Q_{\gamma} \propto t^{q}$, with values up to  $q \sim 4$. This means that even though the electrons are in the fast cooling regime initially (large B field, low injection rate) the dominant contribution to the observed spectra is from emission when the electrons are in a low magnetic field environment, with a correspondingly longer cooling time. The narrowness of the spectrum forces the synchrotron fits, from this scenario, to largely resemble the slow-cooled synchrotron spectrum of \citet{Burgess2014a}.}

 \subsection{Forward Shock Origin}
 \label{alter_scenario}
% see further Burgess et al.
 
% {  extended shock region for he particle acceleratio????} 

 In a baryonic outflow the kinetic energy is typically assumed to be extracted by internal shocks. However, emission from external, forward shocks could be important during the prompt phase for smooth pulsed GRB \citep{Panaitescu1998,Burgess&Begue2015}. %The external shock is created by a spherically expanding blast wave, that is decelerated and energised by sweeping up the circumstellar medium.  
 Similary, \cite{Duffell2014} finds from simulating the jet ($\Gamma \geq 100$) passing though the progenitor star, that a baryon loaded shell lies in front of the jet head at breakout moving with $\Gamma_{\rm shell} \sim 10$. %After a time $t$, %\sim \Delta r_{\rm shell} \Gamma_{\rm shell}^2$, 
 As they collide, highly efficient internal shocks are produced.  The internal shocks are then produced until a radius $\sim 10^{16}$ cm.  
%However, in both the above mentioned scenarios, the constraints imposed by the observed synchrotron emission in the spectra also need to be satisfied i.e. either the electrons should cool slowly or the electrons undergoing fast cooling need to be continuously reaccelerated. 
%{  Why would the emission be slow cooled synchrotron? The photospheric emission would not be absorbed by this baryon loaded shell. }
In both cases, the shocks produced results in synchrotron emission with $\gamma_{\rm min} \sim (m_{\rm p}/m_{\rm e}) \, \Gamma$, which now is more consistent with the inferred values from the observations. 
 
In the external shock scenario the evolution observed for the non-thermal component is  independent of that of the thermal component. The inferred initial Lorentz factor ($\Gamma_0$) from the evolution of the synchrotron component, would not be larger than the largest $\Gamma$ value inferred from the thermal component at the photosphere.  The estimation of the outflow parameters at the photosphere at later times during the decay phase of the pulse is not possible, as the total flux corresponding to each time bin is not known explicitly.

The observed variability of the thermal component with time tells us how the central engine varies with time.
Following the arguments in \S \ref{Rph}, we find that during the rising phase of the pulse if $\Gamma$ remains nearly steady, the baryon load of the outflow is increasing with time. As a result we can expect high inertia shells to be ejected by the central engine in the beginning of the burst. This high inertia shells then crash into the external medium and results in the shocks which then produce the observed synchrotron emission.  %
During the decay phase, as luminosity of the burst is decreasing, the shells ejected by the central engine have lower $\Gamma$ and thereby may not catch up with the external, forward shocks that had been produced. Instead, they may be decelerated by the reverse shocks that have been produced.  A more detailed discussion in this scenario is given in \cite{Burgess&Begue2015}.

\subsection{Evolution of $r_0$}
\label{r_0disc}

%Increasing $r_0$ suggests that there are photons being produced in the region above the central engine (which has a typical size of the order of $10^{6} - 10^{7}$ cm). This means, there is efficient photon production and increased dissipation with time within the limits of the thermalisation radius \citep{Vurm2013}. 
%The deduced increasing behaviour of $r_0$ can be speculated to be the result of colliimation shocks that occur within the progenitor envelope as the jet pierces through \citep{Beloborodov2013, Lazzati2009,Mizuta2013}.  
Shear turbulence and oblique shocks within the stellar core  can result in  $r_0$ attaining {  values much larger than the expected size of the central engine (e.g., \cite{Thompson2007, Iyyani2013, Pe'er2015})}.  
Indeed, several hydrodynamical simulations have shown that there are significant collimation shocks produced within the outflow as the jet traverses through and emerges out of the stellar cocoon \citep{Mizuta&Ioka2013,Lopez-Camara2013,Zhang_MacFadyen2003,Lazzati2015}.   It is also interesting to note that the variability time scale that are found in GRB light curves are consistent with the large $r_0$ values derived above. For instance, \cite{Golkhou&Butler2014} find the typical minimum variability time scales, $\Delta t$, to be of the order of a fraction of a second, with a shortest time scale of 10 ms. Such a time scale corresponds to a size of the central engine between $r < c \Delta t _{\rm min} \sim 3 \times 10^8$ cm and $2 \times 10^{10}$ cm.

\cite{Iyyani2013} suggested that a larger outflow velocity (or $\Gamma$) would prevent the formation of such shocks and thus result in smaller values of $r_0$.  Most of the bursts in the sample have $r_0$ that evolve like a pulse with time. Thus, it may be speculated that during the period where $r_0$ increases and shows a negative correlation with $\Gamma$, the position of $r_0$ may be determined by such shocks when the jet is propagating through the progenitor envelope \citep{Beloborodov2013, Lazzati2009,Mizuta2013}.
%the dissipation othe collimation of the jet \citep{Beloborodov2013, Lazzati2009} by the cocoon of the star resulting in oblique shocks \citep{Mizuta2013}.  
Beyond the core radius of the star, the oblique shocks due to the confinement of the jet by the cocoon of the progenitor becomes weak and less efficient. 
As a result, $r_0$ does not increase any further, instead decreases or remains nearly steady. 
%and $r_0 \sim10^{10}$ cm suggests the size of the core of the Wolf- Rayet star. 
{  However, there have been no direct simulation study done to evaluate how $r_0$ evolves during a GRB mainly due to the limitations of numerical simulations on these scales.}

%Another possibility 
It has been suggested by \cite{Ghisellini2007} that shocks are produced in the outflow when it encounters the cocoon material surrounding the progenitor, which is in the way of the jet. As a result, the fireball is reborn at a larger radius, i.e, the surface of the progenitor. If we associate $r_0$ to the surface of the progenitor, in such a case it may be speculated that with time as the stellar material surrounding the black hole gets accreted, we would expect $r_0$ to decrease. In accordance to this, we find that after $r_0$ reaches its peak value, it then decreases and thereby shows a positive correlation with $\Gamma$. 

%However, we find $r_0$ to increase with time and the values lie between the limits of the size of the event horizon of the black hole formed and the size of the progenitor star. This probably suggests that the radius at which the outflow starts to expand freely is mainly determined by the collimation or oblique shocks created within the outflow when the jet is propagating through the progenitor envelope. But we would like to note that these ideas are mere speculations and till now there have been no direct simulation study done to evaluate how $r_0$ evolves with time. However, several hydrodynamical simulations have shown that there is significant collimation shocks produced within the outflow as the jet traverses through and emerges out of the stellar cocoon.    

%\subsubsection{Correlation between the flow nozzle and its speed}
 %However after reaching the stellar core one can speculate that due to the accretion of the stellar material, $r_0$ starts decreasing with time and thereby shows a positive correlation with $\Gamma$. But 
 
 Yet another alternative explanation to the observed temporal behaviour of $r_0$ is
propagation effects of the jet inside the collapsing star. The jet
expansion is actually not expected to be free, but is affected by various
effects such as multiple recollimation shocks, mass entrainment, or a non-
conical structure of the flow. In such a case, the interpretation of the value of $r_0$  as well as 
its evolution will have another meaning.

\subsection{High latitude effects on the evolution of temperature and flux}

It was proposed by \citet{Pe'er2008,Pe'er&Ryde2011} that the origin of
the late time decay of the thermal flux and the temperature \citep{Ryde2004,Ryde2005, Ryde&Pe'er2009}, may be associated with off-axis emission, which is
seen at a delay with respect to the emission from the jet
axis. Furthermore, the late time photons are observed at lower
energies due to the lower Doppler boost and larger photospheric radius
(which is angle-dependent). However, as was pointed out by \cite{Deng2014}, for spherical outflows with parameters characterising
GRBs, the characteristic time-scale for the decay is faster than
observed here. {  This issue may be resolved if one introduces angular structure of energy and Lorentz
factor of the jet} (Beloborodov 2010, \citet{Lundman2013}), although detailed calculations and
hence firm conclusions are still lacking.

\subsection{Poynting-flux dominated outflows}
\label{sec:Poynting}

{  Even though a strong photospheric emission component disfavours Poynting-flux dominated flows, since the energy of such a component would 
be suppressed by a factor of $(1+\sigma)^{-1}$ \citep{Zhang&Pe'er2009}, one cannot rule out such a possibility. If, e.g., the radiative efficiency is low (i.e large $Y$ parameter) the actual ratio of the thermal to kinetic energy might be smaller than estimated from the data.  {  Furthermore, in Poynting-flux dominated flows many of the problems faced by synchrotron emission in baryonic flows,  and discussed in \S \ref{sec:alt}, can naturally be overcome. As pointed out by \cite{Zhang&Yan2011} a balance between heating and cooling is expected to be established by second-order stochastic acceleration in the turbulent region of the ICMART scenario {  (Internal-Collision-Induced Magnetic Reconnection and Turbulence)}. Moreover, the number of baryons associated electrons is smaller by a factor of $\sim (1+\sigma)^{-1}$, and therefore every electron naturally attains a higher Lorentz factor. %However, the Poynting-flux scenario still needs to account for the strong blackbody components that are observed, as mentioned in \S \ref{r_0r_s}.}%\ref{sec:alt}.
}

In order to estimate the flow parameters and their evolution, as done above for the baryonic dominated case, a more generalised formalism is needed. This is because, if the Poynting flux dominates the energy of the flow, the dynamics will change. For instance, the large fraction of the acceleration phase will have  a more gradual acceleration compared to the initial, thermal acceleration. Moreover, the photosphere will most likely occur while the flow is still accelerating. The general formalism introduced by Gao \& Zhang (2015) covers all these different possible cases. Additional assumptions have to be made, though, in order to estimate the flow properties. First,  further unknowns are introduced (e.g., the magnetisation of the flow $\sigma$). Second, since the acceleration of the flow is assumed to occur in phases with different radial dependencies and since these dependencies are not yet fully understood (see, e.g., \cite{Bromberg&Tchekhovskoy2015}) an assumed prescription is needed to be made.   

Gao \& Zhang (2015) applied this formalism to GRB110721A, whose thermal flux component is only at a few per cent level (see Fig. \ref{fig:ratio}), which is consistent with a Poynting-flux dominated interpretation. They make the assumption of a constant value for $r_0$ (choosing several different values), which allows them to use the observables to derive the flow parameters. With the choice of $r_0 \sim 10^8$ cm, they find that the magnetisation % $\sigma_0$,  
has to vary in strength by nearly 2 orders of magnitude, with the flow changing from being highly magnetised at the photosphere, ($1+\sigma_{  ph}) \sim 100$, to being depleted of the magnetic field (kinetic energy dominates the flow), $(1+\sigma_{  ph}) \sim 1$, at around 2.5 s. At this point, the initial magnetisation is also found to be weak  $(1+\sigma_{0}) \sim 1.5$, which indicates that the acceleration is mainly thermal and that the behaviour approaches that of a baryonic flow. This is consistent with the fact that $r_0$ in Fig. 3 (assuming a baryonic flow) approaches this assumed value of $\sim 10^8$ cm.

%In this discussion, it is relevant to note that the thermal component is still relatively weak when the flow is at its lowest magetisation (around 2.5 s). Then the flux ratio is $F_{\rm BB}/F_{tot} \sim 5\% - 10\%$ (the exact value is dependant on if a Band function or a synchrotron function is used). This flux ratio is significantly  lower than what is found for the other bursts in our analysed sample, which strengthens the assertion that a baryonic approach is a reasonable assumption for these cases.  

%%%%%%%%%%%%%%%%
%Furthermore, in analysing GRB110721A, Gao \& Zhang (2015) find a different evolution of the derived quantity $\Gamma(t)$, as compared to the purely baryonic case. Instead, of finding a monotonic decay \citep{Iyyani2013}  they find an initial phase (until 2.5 s) of weakly increasing (or nearly constant) $\Gamma$, followed by a phase of decaying $\Gamma$, similar to the baryonic case.
%%%%%%%%%%%%%%%%

%Furthermore, in this scenario the non-thermal component cannot be formed in internal shocks.

There are thus two interpretations for the spectral evolution in GRB110721A, a baryonic-dominated flow (this paper and \cite{Iyyani2013}) and a Poynting-flux dominated flow \citep{Gao&Zhang2015}. %The analysis presented here, as well as in Iyyani et al. (2014), assumes a baryonic dominated outflow and can thereby give estimates on, among other poperties, the valus of $r_0$. 
The former interpretation yields a varying $r_0$, which is interpreted as being caused by recollimation shocks as the jet traverses within the star, see \S \ref{r_0disc}.   
%However,  details of the behaviour of such shocks within the star are not yet fully explored and further 
%numerical simulations of hydrodynamical shocks are needed (see, for instance, \cite{Mizuta&Ioka2013, Lopez-Camara2013, Duffell&MacFadyen2014}).
In the latter interpretation, the flow is Poynting flux dominated and the spectral evolution is explained by a varying magnetisation, while keeping $r_0$ fixed. The theory of magnetised outflow, however, is not fully developed. For example, it is not obvious as to which value of $r_0$ should be chosen.
%Here an explanation to why the magnetisation varies as is deduced, why $r_0 \sim 10^8$ cm, 
%why is goes from Poynting to baryonic (the interpretaions coincide). In addition, why are some bursts magnetic and others Baryonic. Developpe Poynting theory.

{   The fraction of bursts in which a strong and/or statistically significant thermal component exists is debatable. In the strongest, single pulsed bursts studied above, 4 out of 8 have such a strong thermal component, such that the magnetic content of the jet cannot be dominant. }
%In order to test between these interpretations, the relative strength of the thermal component is essential, since it provides constraints on the magnetisation. For such a comparison observations are needed for which the parameters $Y$ and $z$ can be determined. 
%%%%%%%%%%%%%%
%Moreover, the determination of the BB is needed to be quantified.
%%%%%%%%%%%%%%%%
On the other hand, the four bursts, in which there are no strong detections of blackbodies, are consistent with Poynting flux-dominated outflows, yeilding a synchrotron spectrum. We point out that in similar cases to these bursts, the width of the spectrum is essential to estimate, since there is a hard limit to how narrow a synchrotron spectrum can be, see \S 5.4 and \cite{Axelsson2015, Yu2015}.

%%%%%%%%%%%%
%Finally, we note that there is a growing consensus on that $r_0$ does not necessarily need to be directly associated with the size of the central engine at $\sim 10^7$ cm, but it can be at a much larger distance of $10^8 -- 10^9$ cm  (Thomson et al. 2006, Iyyani et al. 2013, Ghirlandna et al. 2014, Zhang et al. 2015, Pe'er et al. 2015). 
%%%%%%%%%%%
}

\subsection{Subphotospheric dissipation}
\label{sec:alternative}

This study has interpreted the prompt GRB spectrum in a two-emission-zone model where the blackbody component is from the photosphere and the synchrotron component is from the optically-thin region (see also \cite{Ryde2005, Guiriec2013, Iyyani2013, Burgess2014a, Preece2014}). Such an interpretation describes a totally different physical scenario compared to the one invoked by a fitting model in which the full spectrum is due to subphotospheric dissipation (\cite{Ryde2010,Iyyani2015,Ahlgren2015}). In the latter scenario, all emission stems from the photosphere which no longer forms a blackbody. Due to dissipation of the kinetic energy of the flow in a region below the photosphere the emission spectrum can be significantly broader and have complex shapes \citep{Pe'er&Waxman2005,Beloborodov2010}. 

\begin{figure}
\begin{center}
\resizebox{84mm}{!}{\includegraphics{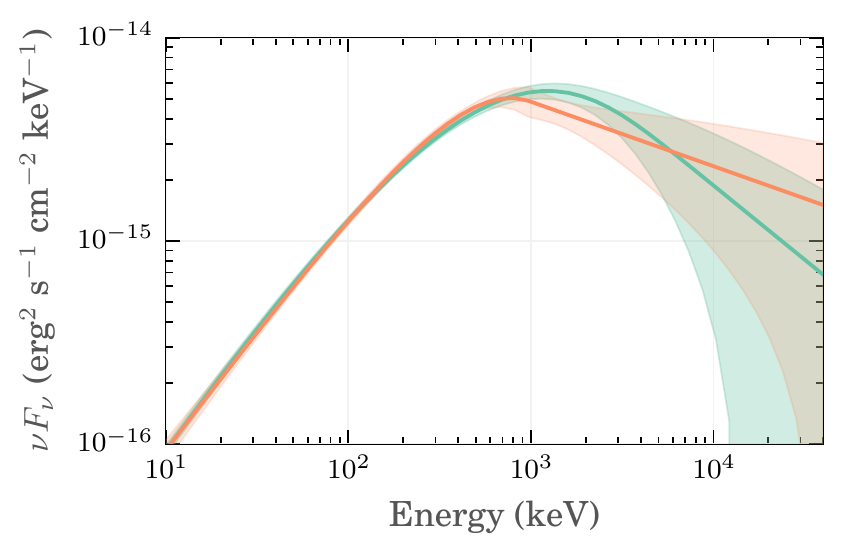}}
\caption{The spectral fits to the time-bin at the peak flux of GRB081110A:  The Band function fit is shown by the orange solid line and the synchrotron fit is  shown by  the green solid line. The shaded regions depict the uncertainty in the spectral shapes of the respective model. The Band function results in a spectral shape that is narrower at the $\nu F_{\nu}$ peak as compared to the shape of  the synchrotron function.}
\label{081110A}
\end{center}
\end{figure}

In many cases, both the models are consistent with the data. This ambiguity in the interpretation of the data is illustrated here for the case of bursts GRB081110A and GRB110920A. GRB081110A is one of the bursts in the sample that is consistent with the synchrotron emission alone, while GRB110920A shows evidence for an additional blackbody component. In Figure \ref{081110A}, we show the spectral fit to the time bin at its peak flux, when modelled using a Band function alone (orange solid line) and a synchrotron function alone (green solid line), where the shaded region shows the uncertainty in the shape of the spectrum related to each model respectively. It is interesting to note that due to the large flexibility of the Band function, the Band function fit to the spectrum results in a spectral shape that is narrower at full width half maximum (FWHM) of the $\nu F_{\nu}$ peak, than the synchrotron function. Such a spectrum is more easily interpreted in a subphotospheric dissipation scenario, which is not limited by the fundamentally required  width for synchrotron emission.
However, we find that statistically both the models are  consistent with the data. In such cases, it becomes important to explore the implications of each model within its related physical scenario which need to be validated if the resultant physical conditions are feasible or not. 
%%%%%%%%
%%%%%%%%
%%%%%%%%
%The Figure \ref{081110A} thus also cautions our interpretation that a narrow spectrum (refers to the spectral width at FWHM of the $\nu F_{\nu}$ peak of the Band function fit) is inconsistent with synchrotron emission \citep{Axelsson2015}.  
%%%%%%%%%%
%%%%%%%%%%
%%%%%%%%%%

\begin{figure}
\begin{center}
\resizebox{84mm}{!}{\includegraphics{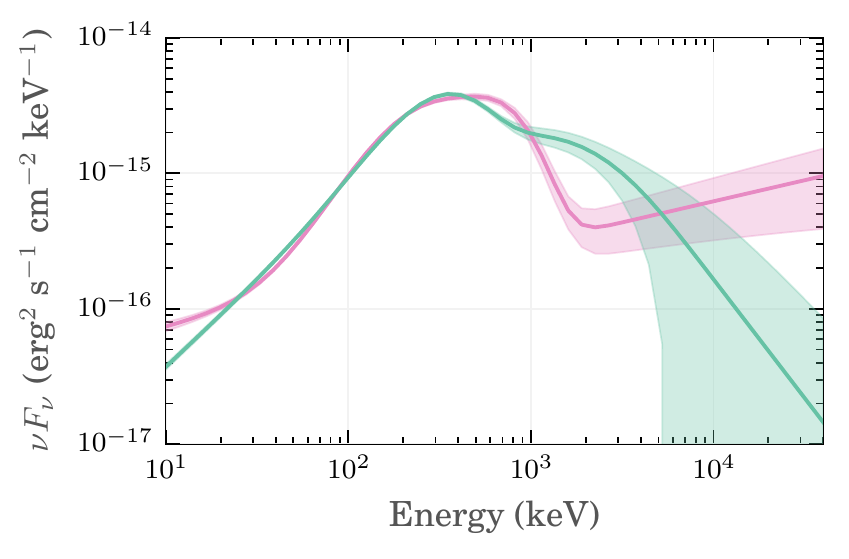}}
\caption{The spectral fits to the time-bin at the peak flux of GRB110920A: Comptonisation + power law model of Iyyani et al. (2015) is shown by the pink solid line and the blackbody+synchrotron model is shown by the green solid line. The spectral shapes resulting from the two models are clearly different. The shaded regions depict the uncertainty in the spectral shapes of the respective model. }
\label{110920A}
\end{center}
\end{figure}

Further, elaborating this fact of complication involved in the interpretation of various models fitted to the spectrum, in Figure \ref{110920A}, we show the comparison of the spectral fits to the spectrum of GRB110920A with the models: Comptonisation + power law (pink solid line), which represents the photospheric emission including localised subphotospheric dissipation at moderate optical depths \citep{Iyyani2015}, and blackbody + synchrotron emission (green solid line), which represents the two-emission-zone model.
% where the blackbody and synchrotron component are associated to the photospheric emission and optically thin non-thermal emission respectively.
 Figure \ref{110920A} clearly shows that these two models result in two different spectral shapes, however, both the spectral shapes are consistent with the data. We note that, in this particular case, we find the narrowness of the spectrum at the $\nu F_{\nu}$ peak is highly constraining, since in blackbody + synchrotron fit, the spectral width at FWHM is given by the blackbody component, which thus confirms the fact that the $\nu F_{\nu}$ peak of the spectrum is actually very narrow. This, again points out the fact that a spectrum which has a narrow $\nu F_{\nu}$ peak, still may be consistent with a synchrotron emission, however, only with a dominant blackbody emission which is observed in this case. We also note that from a statistical (C-stat) point of view, we find Comptonisation + power law to be a better model. However, since both the models are differently motivated, the statistical comparison is largely indecisive. The acquired observational behaviour needed to be physically validated, in order to clearly ascertain which is the best spectral model.  This in turn requires the existing GRB theories to have clear predictions in regard to the temporal behaviour of the outflow in their respective physical picture. 
%is highlighted by GRB110920A in our sample above. In \cite{Iyyani2015}, this particular burst was modelled with a dominant photospheric emission which was formed by Comptonisation at a moderate optical depth, while in the modelling above the photospheric component is a blackbody and is less dominant. In this particular case the Comptonisation model yields a lower C-stat value but the statistical comparison is largely indecisive.  Moreover, {  figure}  shows that even for the burst for which the data is consistent with a single synchrotron spectrum, a Band function gives a similarly good fit. Due to the large flexibility of the Band function the spectrum has a different shape, in particular with a narrower peak. Such a spectrum is more easily interpreted in a subphotosphric scenario which is not limited by the fundamentaly required  width for synchrotron emission.

Fitting physical models is an important step forward in comparison to empirical model fitting which is the most common approach today. However, due to the ambiguity between models we have to resort to assessing the models from a theoretical perspective by considering the constraints that observations set. For instance, in the synchrotron fits above, the data strongly disfavours a typical fast cooling scenario. Slow cooling is permitted but sets strong constraints on the acceleration process needed such that only a small fraction of the electrons are accelerated to very high energies. Moreover, the emission would be greatly inefficient.  This points towards reheating of the electrons causing a steady state electron distribution which produces a slow-cooling-like spectrum. %Such emission should rather be denoted as non-cooled or partially cooled synchrotron emission since even though the cooling happens, very rapid reheating balances it. 
The physical scenario needs to be addressed in order to assess the validity of the fit results.

Another weakness of the two-zone scenario presented above is the use of a blackbody for the photospheric component. The observed emission from the photosphere is expected be a multicolour blackbody  \citep{Beloborodov2010}. Along with that, including the effects of a GRB jet observed at different viewing angles can give rise to a much broader spectrum, see \cite{Lundman2013}. Moreover, there can also be dissipation of the kinetic energy of the flow below the photosphere. In such a case, the blackbody will be significantly broadened and the physical validity of the blackbody+synchrotron model will fall.  However, there is a small but significant fraction of bursts that are indeed fitted with a single blackbody \citep{Ryde2004, Ghirlanda2013, Larsson2015} which suggests that the emission from the photosphere under certain circumstances can be a blackbody. Moreover, the blackbody component used in the fits does not need to reflect the true shape but rather capture a peaked spectrum appearing above the synchrotron emission. 
%The physical requirements and their validity should be further addressed.

%On the other hand, the arguments for broadening the blackbody emission from the photosphere, forms the basis for the subphotospheric dissipation model.  The main concern in such an interpretation, though, is that there is no prescription as to why the dissipation should be local and should occur at moderate optical depths of around a few tens.

%\subsection{Constraining photospheric emission}
%
%Within the sample, there are 3 bursts: GRB081110A, GRB090809A and GRB110407A, whose spectra are consistent with a synchrotron component alone. Since these spectra are consistent with the synchrotron emission from the same electron distribution that has been considered for the other bursts, it is possible for us to assume that the magnetic field, $B$ and electron Lorentz factor, $\gamma_{\rm el}$, have the same constraints as that in those bursts. Considering those constrains enables us to constrain the Lorentz factor and the photospheric radius of the outflow of these bursts, see Figure \ref{fig:gamma_rph_3bursts}.
%We find that the constraints enclose the parameter space that are consistent with the generally observed values. 
%%However, it is worth noting that, in case of GRB090809A, we find the Lorentz factor to have values less than $100$ towards the end of the burst, which results in photospheric radius larger than $10^{13} \rm cm$. 
%
% \begin{figure*}
%\begin{center}
%\resizebox{84mm}{!}{\includegraphics{gamma_rph_081110A_1.eps}}
%\resizebox{84mm}{!}{\includegraphics{gamma_rph_090809A_1.eps}}
%\resizebox{84mm}{!}{\includegraphics{gamma_rph_110407A_1.eps}}
%\caption{The Lorentz factor, $\Gamma$, (grey shaded region) and photospheric radius, $r_{\rm ph}$, (blue barred region) values constrained using the parameter space of estimate of $B$ and $\gamma_{\rm el}$, for bursts: GRB081110A (upper left), GRB090809A (upper right) and GRB110407A (lower middle) are shown.   }
%\label{fig:gamma_rph_3bursts}
%\end{center}
%\end{figure*} 
%

\section{Conclusions}

We have investigated the fits to the GRB spectral data of a sample of 8 single pulsed GRBs, with models involving synchrotron emission. Two immediate requirements apply: (i) {   a photospheric component (blackbody) is strong in four out of eight bursts and statistically highly significant, while subdominant}, in another one and (ii) the energy distribution of the radiating electrons have to be in {  the slow cooling regime or be reheated}.  {  The need for a strong blackbody {suggests} that the flow is baryonic-dominated, {  at least in four of the cases}. Furthermore, we find a robust trend that the Lorentz factor of the flow decreases with time during the burst which can neither be explained by a varying radiative efficiency nor a varying magnetisation of the jet making the reasonable assumption that the photosphere radius is above the coasting radius).} Such a behaviour is contradictory to what is expected from various internal shock scenarios. We also find a strong trend that the distance from the central engine to the flow nozzle can attain large values (typically $10^8 - 10^9$ cm) and increase with time. This gives a prediction to the properties of the jet at the very core of the progenitor star. 

%{  XX discuss fraction with BB comonsnt flux depedence}

The non-cooled electron distribution, required from the fits, can be due to the cooling time being larger than the typical dynamical time (slow cooling). In such a case only a small fraction of the electrons should be accelerated to very large Lorentz factors {  or the energy dissipation should be due to forward shocks}. The radiative efficiency would by necessity then be low in these cases.  Alternatively, the electron distribution can be attained through a balance between heating and cooling. However, this is typically not expected in baryonic shocks, even though a scenario of extended shocks have been described, and therefore, needs further investigation to verify its plausibility. {  Finally, a scenario where the plasma is inhomogenous  such that the magnetic field strength varies have been suggested. Further fits to data with such physical models are needed to assess their validity.
% suggest that most of the electrons emit in low magnetic field regions. 
}

We point out that alternative physical models, for which the synchrotron component is not needed at all, are also consistent with the data. In particular, scenarios including subphotospheric dissipation afford to give physically plausible explanations of the data. However, such models need to explain the inferred radial distribution of dissipation, for instance, through numerical simulations. The fits cannot distinguish between these models decisively from a statistical point of view and the distinction must be made using arguments regarding their physical plausibility. Finally, we have also pointed out that in order to test the viability of synchrotron emission in the observed data, a Band function is not sufficient. \\

\section*{Acknowledgements}

 We thank the referee, Dr. Bing Zhang, for very thorough and useful comments on the manuscript. We acknowledge support from the Swedish National Space Board and Swedish Research Council (Vetenskapsr{\aa}det). SI is supported by the Erasmus Mundus Joint Doctorate Program by Grant Number 2011-1640 from the EACEA of the European Commission.  DB is supported by a grant from Stiftelsen Olle Engkvist Byggm\"astare.   AP acknowledges support by the European Union Seventh Framework Programme (FP7/2007-2013) under grant agreement no 618499.

\bibliographystyle{mn2e}   
 \bibliography{ref}  

\end{document}